\documentclass[12pt]{article}
\usepackage[margin=0.9in]{geometry}
\usepackage[utf8]{inputenc}
\usepackage{bm}
\usepackage{stmaryrd}
\usepackage{amsmath}
\usepackage{amsthm}
\usepackage{amssymb}
\usepackage{xcolor}
\usepackage{graphicx}
\usepackage{ulem}
\usepackage{url}
\usepackage{comment}

\newcommand{\tr}{\mathrm{Tr}}

\newcommand{\h}{{\mathcal H}}

\renewcommand{\r}{{\rm R}}
\newcommand{\oV}{{\overline V}}

\newcommand{\s}{{\rm S}}

\newcommand{\bbbone}{\mathchoice {\rm 1\mskip-4mu l} {\rm 1\mskip-4mu l}
{\rm 1\mskip-4.5mu l} {\rm 1\mskip-5mu l}}
\newtheorem{thm}{Theorem}
\newtheorem{prop}{Proposition}
\newtheorem*{prop*}{Proposition}

\usepackage{authblk}
\title{Quantum systems coupled to environments \\
via mean field interactions}
\author[1]{Michele Fantechi\footnote{michele.fantechi@gmail.com}}
\author[2]{Marco Merkli\footnote{merkli@mun.ca}}
\affil[1]{Dipartimento di Matematica

Politecnico di Milano 

P.zza Leonardo da Vinci, 32, 20133, Milano, Italy
\medskip
}
\affil[2]{Department of Mathematics and Statistics

Memorial University of Newfoundland

St.~John’s, NL, Canada A1C 5S7
}

\begin{document}

\maketitle
\begin{abstract}
We show that when a quantum system is coupled to an environment in a mean field way, then its effective dynamics is governed by a unitary group with a time-dependent Hamiltonian. The time-dependent modification of the bare system Hamiltonian is given by an explicit term involving the reservoir state. We show that entanglement within the system state is not changed during the dynamics. Our results hold for arbitrary strengths of the system-environment coupling, and for finite or infinite dimensional systems. As an application we show that the qualitative dynamical features of an $N$-body system can be altered drastically by the contact with the environment. For instance, bound states can turn into scattering states and vice-versa.  
\end{abstract}

\section{Introduction}

Open quantum systems are often described as a bipartite complex of a quantum system $\s$ in contact with a quantum environment, also called a reservoir $\r$. The system typically consists of one or several quantum particles, spins or qubits and the bath is commonly modeled by a collection of oscillatory degrees of freedom, such as a collection of quantum harmonic oscillators or a quantum field. The interaction between $\s$ and $\r$ is specified by an interaction operator in the total Hamiltonian governing the unitary dynamics of the {\it entire} complex $\s+\r$.  The reservoir is typically considered very `large' when compared to the system. This can be expressed for instance in terms of a sizeable spatial extension, a high density of the energy spectrum, or a large number of particles or other degrees of freedom in the reservoir. The influence of the reservoir on the system is often considered as `noise' and one is interested in the resulting {\it effective} dynamics of the system alone, that is, an effective evolution equation for the system state. The latter must incorporate effects of the reservoir and it carries information about the state of the reservoir -- such as its temperature in an equilibrium situation. 

\begin{figure}[h!]
\centering
\includegraphics[width=.9\textwidth]{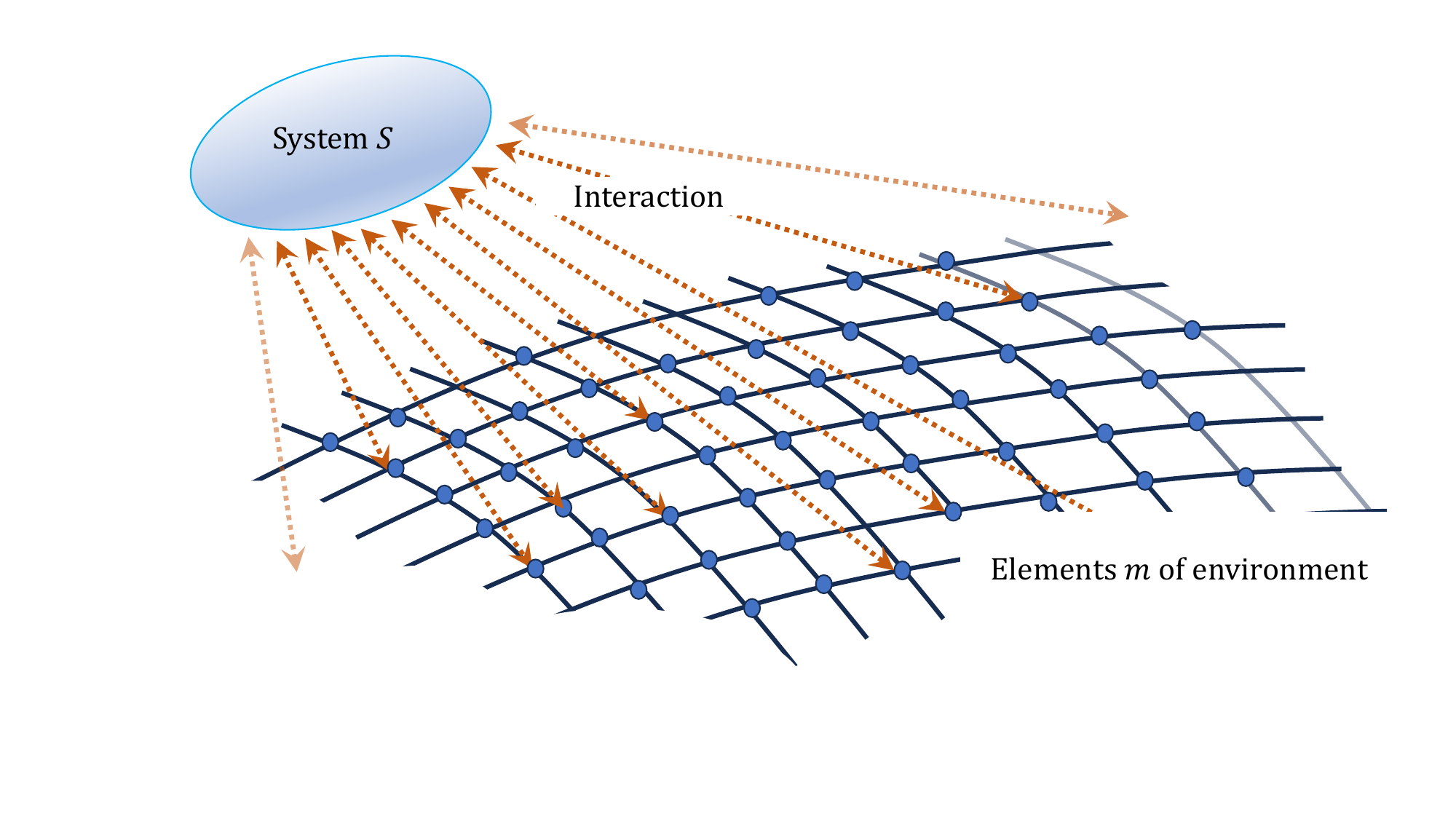}
\vspace*{-1.5cm}
\caption{The system $\s$ interacts with elements $m=1,2,\ldots,M\rightarrow\infty$ of the environment in a uniform way, the same for all $m$. In this rendition, $m$ is interpreted as the label indicating the spatial location of a subsystem of the reservoir. An example for the reservoir is a  lattice spin system, where a spin is located at each position $m$ on a lattice.}
\label{figure1} 
\end{figure}

Finding the exact effective system dynamics is generally impossible except for very simple models. The {\it Markovian Master Equation} is a very popular approximation which can be justified under suitable assumptions. It is a relatively simple equation for the system state which approximates the true system dynamics well, for instance, if the $\s\r$ interaction is weak and if the reservoir loses its memory quickly - notions which can be made physically and mathematically rigorous \cite{BP,RH,Da1, Da2, MAoP, MMQuantum1, MMQuantum2, MMAHP}. In the markovian approximation, the state (density matrix) of a system at time $t$ is obtained by applying a dynamical semigroup (in $t$) to the initial system state. The mathematical structure of this approximate dynamics is simple, entirely encoded in the properties of the generator of the semigroup, and thus amenable to analytic and numerical analysis. 
It is routinely used to examine irreversible phenomena of open system dynamics, such as the approach of a stationary state (thermal equilibrium or non-equilibrium stationary states), as well as decoherence and the evolution of the entanglement between subparts of $\s$. If the $\s\r$ coupling is not small, though, or if the reservoir does not have a short memory, then the markovian approximation deviates substantially from the true dynamics. In particular, it obviously cannot capture any non-Markovian effects. In the {\it ultra-strong} coupling limit, where the $\s\r$ interaction term dominates by far the system energy scales, some recent progress on the effective system evolution has been established with theoretical  and numerical methods \cite{Trush,Janet,Anto1,Anto2,Rivas}. However, an approach which covers all scales of $\s\r$ interaction strengths is not available \cite{Trush+}. In the current work we derive the effective system dynamics mathematically rigorously and without making any smallness assumption on the system-reservoir coupling, thus {\it covering the whole range, from weak to strong to ultra-strong couplings}. We are able to do this for a class of open systems in which the coupling to the reservoir is of {\it mean field} nature.

\medskip

\begin{figure}[h!]
\centering
\includegraphics[width=.9\textwidth]{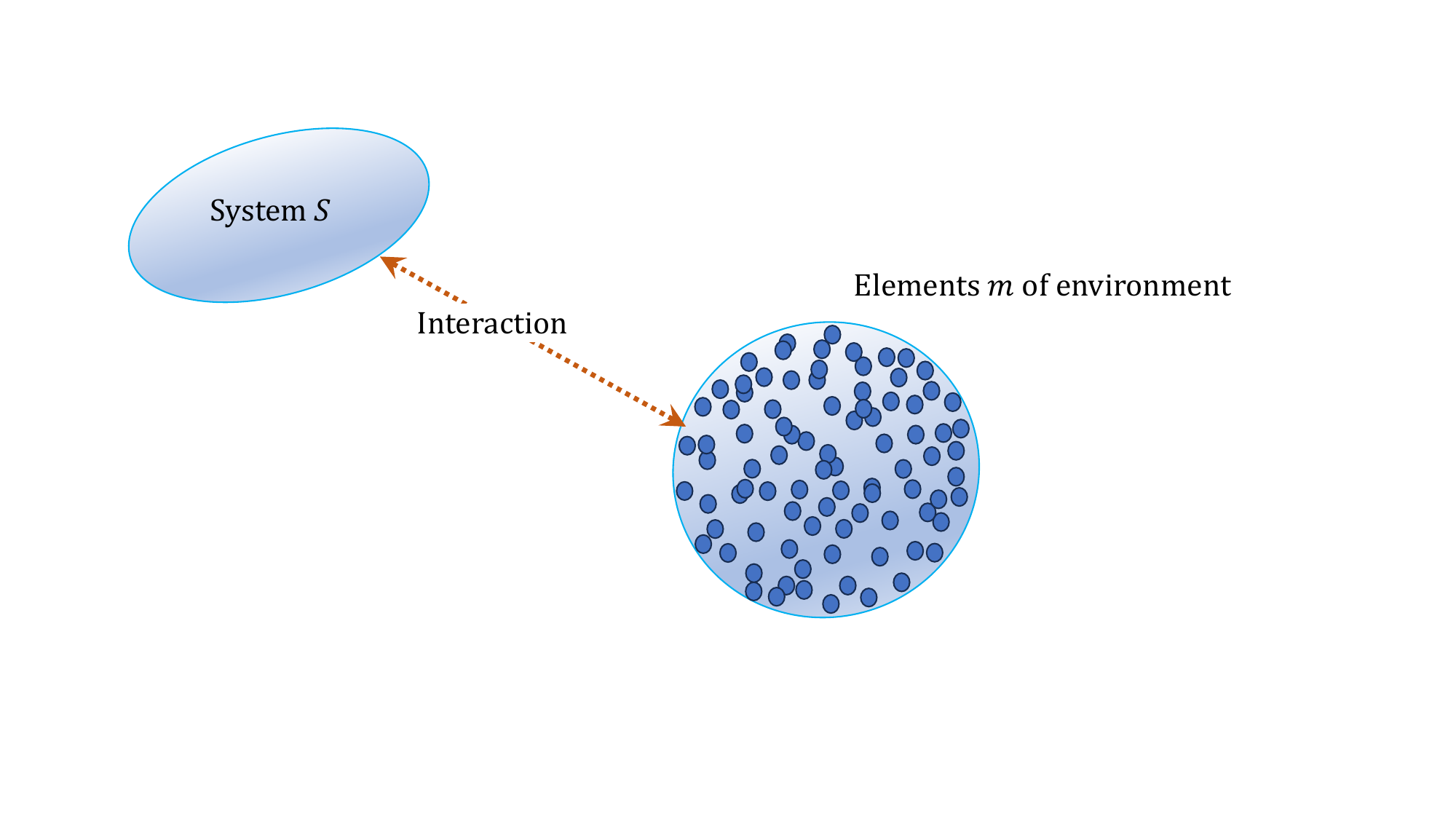}
\vspace*{-1.5cm}
\caption{The system $S$ interacts with elements $m=1,2,\ldots,M\rightarrow\infty$ of the environment in a uniform way. In this rendition, $m$ is labeling the units of a `densely packed' environment. For instance, $m$ may label the modes of oscillatory degrees of environment which may all lie spatially close together.}
\label{figure2} 
\end{figure}

Figures \ref{figure1} and \ref{figure2} give pictorial representations of the setup. We assume that $\r$ consists of a very large number $M$ of identical subsystems, and that $\s$ is interacting with each reservoir subsystem in the same way. Each individual interaction term is scaled by a factor $1/M$, resulting in an $\s\r$ interaction having the same order of magnitude as the bare system Hamiltonian $H_\s$ (no interaction). The symmetry of the interaction operator together with the mentioned scaling is the trademark of  {\it mean field} theories \cite{BPS,Golse+}.

\bigskip

\begin{figure}[h!]
	\centering
	\includegraphics[width=.9\textwidth]{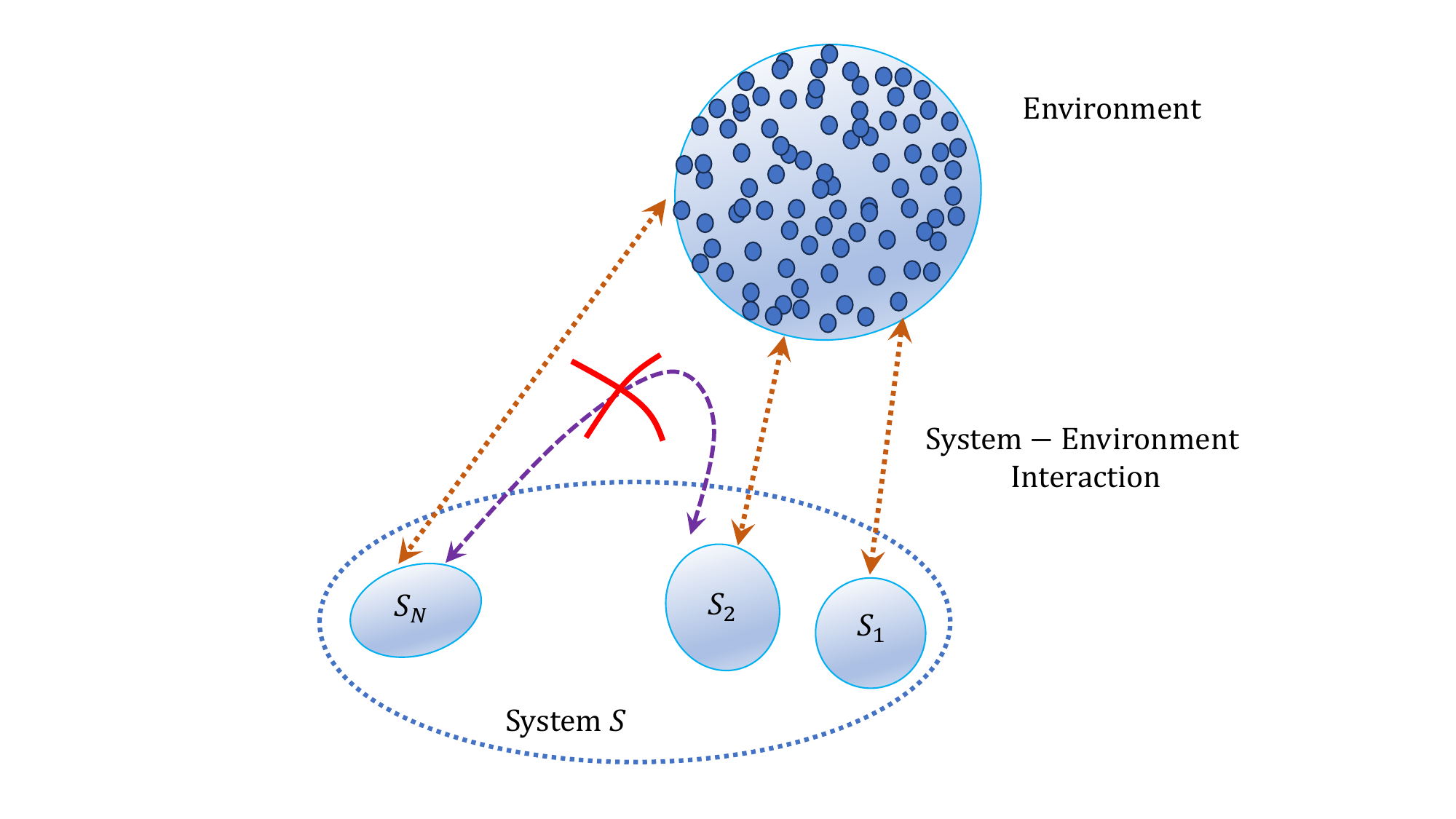}
	\caption{The system $\s$ consists of $N$ (not necessarily identical) subsystems $\s_1,\ldots, \s_N$, each of which interacts with the environment separately. Theorem \ref{thm2} shows that  the dynamics of $\s_1,\ldots,\s_N$ do not get coupled via the indirect interaction with the common reservoir. The mean-field type coupling does not change the entanglement among subparts of the system during the course of the dynamics. The entanglement is not degraded, it is protected.}
	\label{figure3} 
\end{figure}

\noindent
Our {\bf main results} are: 
\begin{itemize}
\item[1.] In the limit $M\rightarrow\infty$ the system evolves according to the Schr\"odinger equation with a generally time-dependent Hamiltonian $H_{\s,\rm eff}(t) = H_\s +W(t)$. The latter is obtained from the bare system Hamiltonian $H_\s$ by adding an explicit time-dependent `potential' $W(t)$. The result is not perturbative, it holds for any $\s\r$ interaction strength. This is the content of our Theorem \ref{thm1} below. 

\item[2.]  The $\s\r$ interaction does not affect the entanglement between the subparts of $\s$. Consider a system $\s=\s_1+\cdots+\s_N$ composed of $N$ subparts (which do not have to be of the same nature), where each subpart is coupled to $\r$ in a mean-field manner as described above, but the subparts are not interacting directly. Then the entaglement within the initial system state does not change in time even though the subparts are all coupled to the same environment. This is the content of our Theorem \ref{thm2}, which is illustrated in a picture in Figure \ref{figure3}.   
\end{itemize}

\medskip

{\bf Outline of the paper. } In Chapter \ref{sec:mainres} we define the class of models and the main object of interest $\rho_\s(t)$, the reduced system state at time $t$. We present and discuss our main results in Theorems \ref{thm1} and \ref{thm2}. In the last section of the chapter we explain how the results apply to reservoirs with correlations. Chapter \ref{sec:appli} contains concrete examples of systems, reservoirs and their interactions for which our general results hold. In Chapter \ref{sec:extensions} we present several extensions of Theorem \ref{thm1}. Chapter \ref{sec:proofs} contains proofs of the main theorems and other results presented in in the other parts. 

\medskip

{\bf Acknowledgements. } Both authors thank Michele Correggi and Marco Falconi for lively,  stimulating discussions. MM acknowledges the financial support from a Discovery Grant by the Natural Sciences and Engineering Research Council of Canada (NSERC), as well as the warm hospitality of the Department of Mathematics of the Politecnico di Milano. MF acknowledges the support of the MUR grant “Dipartimento di Eccellenza 2023-2027” of Dipartimento di Matematica,
Politecnico di Milano, and the support of the ``Gruppo Nazionale per la Fisica Matematica"(GNFM) of the Istituto Nazionale di Alta Matematica ``F.
Severi"(INdAM).

\section{Main results}
\label{sec:mainres}

Our two main results are Theorem \ref{thm1}
which shows that the effective system dynamics is unitary with an explicit time-dependent Hamiltonian, and Theorem \ref{thm2} which implies that the intra-system entanglement is constant during the evolution. We present these results in Sections \ref{subs:effdyn} and \ref{subs:protent}, after introducing the model in Section \ref{subs:model}. In Section \ref{subs:incorr} we describe a class of admissible reservoir states with entanglement between different reservoir sites.

\subsection{Model}
\label{subs:model}

The Hilbert space of states of the system-reservoir complex is 
\begin{equation}
\label{1}
\h_{\s\r,M} = \h_\s\otimes\h_{\r,M},\qquad \h_{\r,M} =\h_\r^{\otimes M}\equiv  \h_\r\otimes\cdots \otimes\h_\r,
\end{equation}
where $\h_\s$ is the system Hilbert space (of dimension $\le \infty$) and each of the $M$ factors $\h_\r$ is the Hilbert space (of dimension $\le \infty$) of a single reservoir subsystem. Given an operator $X_\r$ acting on $\h_\r$ we write
\begin{equation*}
X_\r^{[m]} = \bbbone\otimes \cdots \otimes \bbbone\otimes X_\r \otimes\bbbone\otimes\cdots \otimes \bbbone,
\end{equation*}
where $X_\r$ sits on the $m$th factor in the $M$-fold product. The total Hamiltonian is given by
\begin{equation}
\label{3}
H_M = H_\s + H_{\r,M} +  V_M,
\end{equation}
where $H_\s$ is a self-adjoint operator  on $\h_\s$ and where the reservoir Hamiltonian is 
\begin{equation}
	\label{4}
H_{\r,M} = \sum_{m=1}^M h_\r^{[m]},
\end{equation}
with $h_\r$ a self-adjoint operator on $\h_\r$. The interaction operator has the mean-field form
\begin{equation}
\label{M}
V_M = G\otimes \oV_{\r,M},\qquad \oV_{\r,M}=\frac1M \sum_{m=1}^M v^{[m]},
\end{equation}
where $G$ and $v$ are self-adjoint operators on $\h_\s$ and $\h_\r$, respectively. The system is interacting in a mean field way with every one of the $M$ subsystems of the reservoir.  Denote the free Heisenberg dynamics of a system observable $X_\s$ and a (`single site') reservoir observable $X_\r$ as
\begin{equation}
\label{8}
X_\s(t) =e^{i t H_\s} X_\s e^{-i t H_\s},\qquad X_\r(t) =e^{i t h_\r} X_\r e^{-i t h_\r}
\end{equation}
and set 
\begin{equation}
\oV_{\r,M}(t) = \frac1M \sum_{m=1}^M \big(v(t)\big)^{[m]}. 
\end{equation}
We consider initial system-reservoir states of the form
\begin{equation}
\rho_{\s\r,M} = \rho_\s\otimes \rho_{\r,M},
\end{equation}
where $\rho_\s$ is a density matrix of $\h_\s$ and $\rho_{\r,M}$ is a density matrix of $\h_{\r,M}$. A density matrix $\rho$ on either of $\h_\r$, $\h_\s$ or $\h_\r^{\otimes M}$ gives rise to a `state' 
$$
\omega(\cdot ) = {\rm Tr}_\h\big(\rho \cdot\big)
$$
where the trace is taken over the Hilbert space $\h$ on which $\rho$ acts. The $\omega(\cdot)$ is a positive linear functional on observables (operators), normalized as $\omega(\bbbone)=1$. We will use the notion of density matrix or linear functional on observables interchangeably. We consider reservoir states $\omega_{\r,M}$ (equivalently, the density matrices $\rho_{\r,M}$) which have certain symmetry and product properties as $M\rightarrow\infty$. This is central to the mean field approach. The main property we require is that there is a single site reservoir state $\omega_\r$ (acting on operators on $\h_\r$) such that for any integer $n$, any times $t_1,\ldots,t_n\ge 0$ we have
\begin{equation}
\label{20.01}
\lim_{M\rightarrow\infty} \omega_{\r,M} \big(\overline V_{\! M,1}(t_1)\cdots \overline V_{\! M,n}(t_n)\big) = \omega_\r(v(t_1))\cdots \omega_\r(v(t_n)).
\end{equation}
 The easiest state satisfying \eqref{20.01} is an $M$-fold product  
$$
\omega_{\r,M}=\omega_\r^{\otimes M}\equiv \omega_\r\otimes\cdots\otimes \omega_\r
$$ 
of a single site reservoir state $\omega_\r$ on $\h_\r$. However, more complicated $\omega_{\r,M}$ having correlations between different sites still satisfy \eqref{20.01}. We discuss this in Section \ref{subs:incorr} below.

\subsection{Effective system dynamics}
\label{subs:effdyn}

Our goal is to find an evolution equation for the dynamics of $\s$ alone, that is, for the reduced system state coupled to a reservoir in the limit $M\rightarrow\infty$,
\begin{equation}
\label{7}
\rho_{\s}(t)  = \lim_{M\rightarrow\infty} {\rm Tr}_{\r,M} \Big( e^{-i t H_{\s\r,M}} \rho_{\s\r,M} e^{i t H_{\s\r,M}}\Big).
\end{equation}
Here, ${\rm Tr}_{\r,M}$ denotes the partial trace over the Hilbert space $\h_{\r,M}$. We assume ({\it c.f.} \eqref{20.01}) that for any integer $n$, any $t_1,\ldots, t_n\ge 0$,
\begin{equation}
\label{10}
\sup_{M\in\mathbb N} \big| \omega_{\r,M} \big( \oV_{\r,M}(t_1)\cdots \oV_{\r,M}(t_n) \big)\big| \le C_{\r,n}  \, b_\r(t_1)\cdots b_\r(t_n),
\end{equation}
and that (recall \eqref{8})
\begin{equation}
	\label{10.2}
\big \|G(t_{\ell_1})\cdots G({t_{\ell_L}})  \rho_\s G(t_{\ell_{L+1}}) \cdots G(t_{\ell_n})\big\|_1 \le C_{\s,n}  \, b_\s(t_1)\cdots b_\s(t_n).
\end{equation}
Here, $C_{\r,n}$, $C_{\s,n}$ are some constants and $b_\s(t)$, $b_\r(t)$ are some functions. 
In \eqref{10.2}, the $t_{\ell_j}$ are an arbitrary permutation of the $t_1,\ldots, t_n$ and $L=0,\ldots,n$  divides the factors of $G$ into two groups (to the left and the right of $\rho_\s$; $L=0$ corresponds to all the $G$ being to the right of $\rho_\s$, similarly for $L=n$). Also,  $\|\cdot\|_1$ is the trace norm on bounded operators acting on $\h_\s$, that is, for a bounded operator $X$ on $\h_\s$, $\|X\|_1 = {\rm Tr}(\sqrt{X^* X})$. The non-negative functions 
 $b_\s(t)$ and $b_\r(t)$ must be such that for all $t\ge 0$,
\begin{equation}
B(t)\equiv	\int_0^t b_\s(s)b_\r(s) ds <\infty.
\end{equation}
We do allow for $B(t)\rightarrow\infty$ as $t\rightarrow\infty$ (so $b_\s$ and $b_\r$ are very general, it suffices that they be locally integrable). We further assume that  for all $t\ge 0$,
\begin{equation}
\label{cond}
\sum_{n\ge 0} \frac{(2B(t))^n}{n!} C_{\s,n}C_{\r,n} <\infty.
\end{equation}
A simple example where all conditions \eqref{10}-\eqref{cond} are satisfied is when $\|G\|, \|v\|<\infty$ (operator norm of linear operators acting on $\h_\s$ and $\h_\r$, respectively). Then we can take $C_{\s,n} = \|G\|^n$, $C_{\r,n} = \|v\|^n$, $b_\s(t)=b_\r(t)=1$,  $B(t)=t$ and \eqref{cond} is satisfied for all $t\ge 0$. This case of bounded operators includes {\it all} situations where $\dim\h_\s<\infty$ and $\dim\h_\r<\infty$. 

\begin{thm}
\label{thm1}
The system state \eqref{7} is given by
\begin{equation}
\label{12}
\rho_\s(t) = U(t) \rho_\s(0) U(t)^*,
\end{equation}
where $U(t)$ is the unitary on $\h_\s$  solving the equation
\begin{equation}
i\partial_t U(t) = H_{\s,\rm eff}(t) U(t),\quad U(0)=\bbbone 
\label{18}
\end{equation}
with the effective Hamiltonian
\begin{equation}
\label{18.1}
H_{\s,\rm eff}(t) =H_\s + \omega_\r\big(v(t)\big) G.
\end{equation}
\end{thm}
The effective system equation is thus the Schr\"odinger equation with time-dependent Hamiltonian $H_{\s,\rm eff}(t)$. The interaction with the reservoir induces a time-dependent `potential' $\omega_\r(v(t))$. If $\omega_\r$ is stationary (with respect to the single site reservoir dynamics generated by $h_\r$) then the potential is constant in time. We note that the result is not a weak coupling result -- there is no `small coupling constant' entering the $\s\r$ interaction.
\medskip

\noindent
{\bf Remark.} The unitary group generated by a time-dependent generator $H(t)$ is a two-parameter group $U(t,s)$ satisfying $i \partial_t U(t,s) = H(t) U(t,s)$, $U(s,s)=\bbbone$. It obeys the composition rule $U(t,t_0) =U(t,s) U(s,t_0)$. In this paper, we simply write $U(t)=U(t,0)$.

\subsection{Protection of intra-system entanglement}
\label{subs:protent}

Consider the system $\s$ to be made of $N$ subsystems,
\begin{equation}
	\label{m14}
\h_\s = \h_{\s,1}\otimes \cdots \otimes\h_{\s,N}
\end{equation}
with Hamiltonian
\begin{equation}
H_\s = H_{\s,1}+\cdots + H_{\s,N},
\end{equation}
where $H_{\s,j}$ is the Hamiltonian of the $j$th subsystem, acting on $\h_{\s,j}$. We do not assume that the $N$ subsystems are of the same nature. In particular, the dimension of $\h_{\s,j}$ may depend on $j$. We take the interaction between $\s$ and $\r$ to be 
\begin{equation}
V_M = \sum_{j=1}^N G_j\otimes \overline V_{\r,M,j},
\end{equation}
where $G_j$ is a self-adjoint operator on $\h_{\s,j}$ and we also write simply $G_j$ for the operator acting trivially on all factors of $H_\s$ except on the $j$th one. Moreover,  
\begin{equation}
	\label{m17}
	\overline V_{\r,M,j} = \frac{1}{M}\sum_{m=1}^M v_j^{[m]},
\end{equation}
with $v_j^{[m]}\equiv (v_j)^{[m]}$ and $v_j $ an operator on $\h_\r$.  This means that each subsystem $j=1,\ldots,N$ interacts with the reservoir  independently via a mean-field interaction of the type \eqref{M}. The total Hamiltonian is given by
\begin{equation}
\label{m18}
H_{\s\r,M} = H_\s +H_{\r,M}+ V_M,
\end{equation}
where $H_{\r,M}$ is as in \eqref{4}. The reduced system state in the limit $M\rightarrow\infty$ is defined in \eqref{7}.  Analogously to \eqref{10}, \eqref{10.2}, we assume that 
\begin{equation}
	\label{10.1}
\sup_{M\in\mathbb N}\big| \omega_{\r,M} \big( \oV_{\r,M,{j_1}}(t_1)\cdots \oV_{\r,M,{j_n}}(t_n)\big)\big| \le C_{\r,n}\, b_\r(t_1)\cdots b_\r(t_n),
\end{equation}
\begin{equation}
	\label{10.3}
\big \|G_{j_1}(t_{\ell_1})\cdots G_{j_L}({t_{\ell_L}})  \rho_\s G_{j_{L+1}}(t_{\ell_{L+1}}) \cdots G_{j_n}(t_{\ell_n})\big\|_1 \le C_{\s,n}  \, b_\s(t_1)\cdots b_\s(t_n),
\end{equation}
where $\ell_k$ is any permutations of $\{1,\ldots,n\}$ and $j_1,j_2,\ldots,j_n\in\{1,\ldots,n\}$. The constants $C_{\s,n}$, $C_{\r,n}$ and the functions $b_\s(t)$, $b_\r(t)$ are assumed be independent of $j_1,\ldots,j_n$ and to satisfy  \eqref{cond}. Again, an example where the conditions are satisfied is when $\|G_j\|, \|v_j\|<\infty$. Then we can take $C_{\s,n} = \max_{1\le j\le N}\|G_j\|^n$, $C_{\r,n} = \max_{1\le j\le N} \|v_j\|^n$, $b_\s(t)=b_\r(t)=1$ and $B(t)=t$. This includes all situations where $\dim \h_{\s,j}<\infty$, $\dim\h_\r<\infty$.

\begin{thm}
	\label{thm2}
In the setup \eqref{m14}--\eqref{m18}, the system state \eqref{7} is given by
	\begin{equation}
		\label{12.1}
		\rho_\s(t) = U(t) \rho_\s(0) U(t)^*,
	\end{equation}
	where $U(t)$ is the unitary on $\h_\s=\h_{\s,1}\otimes\cdots\otimes \h_{\s,N}$  solving the equation \eqref{18} with the effective Hamiltonian
	\begin{equation}
H_{\s,\rm eff}(t) = H_\s + \sum_{j=1}^N \omega_\r\big(v_j(t)\big) G_j.
		\label{24.1}
	\end{equation}
\end{thm}

Theorem \ref{thm2} shows that the mean-field reservoir does not entangle systems which are indirectly coupled by it, see also Figure \ref{figure3}. It shows that the amount of entanglement within the system is preserved during the dynamics with the reservoir. Indeed, the propagator is a product of local unitaries, $U(t)=U_1(t)\otimes\cdots \otimes U_N(t)$, each given by
$$
i\partial_t U_j(t) =\Big(H_{\s,j}  + \omega_\r\big(v_j(t)\big) G_j\Big) U_j(t),\quad U_j(0)=\bbbone.
$$
Consequently, the amount of entanglement in the system state is independent of time. This is so since a basic property of any measure of entanglement is that entanglement does not change under local unitary operations \cite{PlenioVirmani2007}. In particular,  if $\rho_\s(0) = \rho_{\s_1}\otimes\cdots\otimes \rho_{\s_N}$ is initially of product form, then it will still be so at all later times $t>0$, with independently evolving factors
$$
\rho_\s(t) = \rho_{\s,1}(t)\otimes \cdots \otimes \rho_{\s,N}(t),\qquad 
\rho_{\s,j}(t) = U_j(t) \rho_{\s,j}(0) U_j(t)^*.
$$
Generally, systems coupled indirectly via an environment without mean-field coupling {\it do} get entangled among each other during the dynamics. The creation of entanglement between not directly interacting systems due to a common reservoir was first observed in \cite{Braun} and intricate processes of sudden death and revival of intra-system entanglement in such models was exhibited in \cite{YuEb1, YuEb2, Bellomo, M11}. However, the mean-field nature of the coupling considered here prevents this capacity to generate (or alter in any way) entanglement within the system. In other words, subsystems which are interacting indirectly via a common reservoir by a mean-field coupling do not feel each other's presence - there is no effective subsystem interaction created by the reservoir. This may be seen as a classical feature -- the same phenomenon of absence of the creation of intra-system entanglement occurs indeed when quantum systems are coupled indirectly by the interaction with a reservoir in the classical limit \cite{CFO,CFM,CFFM}, which leads to a similar formal setting. In these references, the particular aspect was not addressed. The connection between mean field and semiclassical theories has been elucidated in  \cite{AmNi1,AmNi2}.

\subsection{Initially correlated reservoir states}
\label{subs:incorr}

The purpose of this section is to exhibit some correlated states satisfying \eqref{20.01}. We start by defining a correlation length $L$ in a  reservoir state. 

\medskip

{\bf Definition (Reservoir correlation length). \ }{\it We say  the state $\omega_{\r,M}(\cdot) ={\rm Tr}_{\r,M} (\rho_{\r,M}\, \cdot)$ on $\h_\r^{\otimes M}$ has correlation length $L\ge 1$ if the following property holds. There is a state $\omega_\r$  on $\h_\r$ such that for any integer $n$ and any operators $x_1,\ldots,x_n$ on $\h_\r$ we have 
\begin{equation}
\label{corrstate}
    \lim_{M \rightarrow \infty} \sup_{\substack{m_1,\dots,m_n =1, \dots ,M \\
    \forall i \neq j, |m_i-m_j|>L }} \big|\omega_{\r,M}  \big( x_1^{[m_1]} \cdots x_n^{[m_n]}\big) -  \omega_{\r} \big(x_1\big)\cdots \omega_{\r} \big(x_n\big)  \big| = 0.
\end{equation}
}
\medskip

The property \eqref{corrstate} is in line with the construction of the `thermodynamic limit' of quasi-local, translation invariant systems. For `local' observables $x_j^{[m_j]}$ (located at fixed `sites' $m_j$) the state is not changing if $M$ is large enough (growing volume), and it factorizes into position independent factors for sufficiently separated observables. 
\medskip

The next result shows that the property \eqref{corrstate} implies the wanted asymptotic factorization, relation \eqref{20.01}.

\begin{prop}
\label{prop1'}
Suppose $\omega_{\r,M}$ satisfies \eqref{corrstate} and let $x_j$, $j\ge 1$, be operators on $\h_\r$ such that all the moments in the state $\omega_{\r,M}$ are finite: For any $n$ there is an $M_0(n)$ and a $C(n)$ such that for $M\ge M_0(n)$,
\begin{equation}
\label{finitemoment}
\omega_{\r,M} \big(x_1^{[m_1]}\cdots x_n^{[m_n]}\big)\le C(n).
\end{equation} 
Define the operators $\overline X_{M,j} = \frac1M \sum_{m=1}^M x_j^{[m]}$ on $\h_{\r,M}$. Then we have, for any $n\in\mathbb N$,
\begin{equation}
\label{20.1}
\lim_{M\rightarrow\infty} \omega_{\r,M}\big(\overline X_{M,1}\cdots \overline X_{M,n}\big) = \omega_\r(x_1)\cdots \omega_\r(x_n).
\end{equation}
\end{prop}
\medskip

\noindent
We present a proof of Proposition \ref{prop1'} in Section \ref{sect:prop1'}.
\medskip

{\it Examples: } 
\begin{itemize}
\item[--] A product state $\omega_{\r,M} = \omega_\r\otimes\cdots\otimes\omega_\r$ has correlation length $L=1$. 

\item[--] Let $E_L$ be a quantum channel acting on $L\ge 2$  subsystems of $\r$, that is, acting on states of the $L$-fold product $\h_\r\otimes\cdots\otimes\h_\r$. For $j=0,\ldots,M-L$, define the state 
\begin{equation}
\label{08}
    \omega_j = \omega_\r^{\otimes j} \otimes E_L( \omega_\r^{\otimes L} ) \otimes \omega_\r^{\otimes M-L-j},
\end{equation}
where $\omega_\r$ is a given state on $\h_\r$. In the state $\omega_j$ the cluster of $L$ subsystems with indices $j+1,\ldots,j+L$ are modified (entangled) by the action of $E_L$, while the remaining subsystems are left in the state $\omega_\r$. By averaging over all positions $j$ we obtain a state 
\begin{equation}
\label{stateRM}
    \omega_{\r,M} =\frac{1}{M-L+1} \sum_{j=0}^{M-L} \omega_j
\end{equation}
which satisfies \eqref{corrstate} with correlation length $L$. To see this, we fix any $m_1,\ldots,m_n$ such that $|m_k-m_l|>L$ for $k\neq l$. Then $\omega_j(x^{[m_1]}_1\cdots x_n^{[m_n]})$ equals $\omega_\r(x_1)\cdots \omega_\r(x_n)$ for {\it most} of the indices $j=0,\ldots, M-L$. Indeed, this is not the case only for values of $j$ such that the set $\{j,j+1,\ldots,j+L\}$, where the operator $E_L$ acts, contains one of the $m_1,\ldots,m_n$. There are $nL$ such values of $j$ in the sum \eqref{stateRM}. Thus $\omega_{\r,M}(x_1^{[m_1]}\cdots x_n^{[m_n]}) = \frac{M-L+1-nL}{M-L+1} \omega_\r(x_1)\cdots \omega_r(x_n)+T(M,n)$, where $|T(M,n)|\le \frac{nL}{M-L+1}C(n)$, provided the $\omega_j$ satisfy the finite moment condition \eqref{finitemoment}.

\item[--] As a concrete example, consider $\h_\r={\mathbb C}^2$ (reservoir made of spins $1/2$) and take $\omega_\r ={\rm Tr} ( |0\rangle\langle 0|\, \cdot) = \langle 0|\cdot|0\rangle$ to be the ground state (relative to $\sigma_z$), with density matrix $|0\rangle\langle 0|$. Then consider a quantum channel $E_2$ which produces a Bell state, $E_2|00\rangle\langle 00| = \frac{1}{2}(|00\rangle+|11\rangle)(\langle 00| +\langle 11|)$. In the corresponding state $\omega_{\r,M}$, \eqref{stateRM}, consecutive elements of the reservoir are maximally entangled.
\end{itemize}

\paragraph{De Finetti type states.} The Quantum De Finetti theorem describes the interplay between symmetry of a state of an infinite system and the absence of entanglement in its finite parts. It was shown in \cite{HuMo} in a form similar to what we discuss here, and later more refined versions were developed in view of applications to mean-field theory, for example in \cite{Rougerie,RoNaLe1,RoNaLe2}. 
\medskip

For $M\ge 1$, let $\omega_{\r,M}$ be a state defined on $\h_\r^{\otimes M}$ given by a density matrix $\rho_{\r,M}$, with the following properties:
\begin{itemize}
    \item[i)] (Consistency) Given any bounded operators $X_1,\ldots,X_M$ on $\h_\r$, 
we have $\omega_{\r,M+1} ( X_1 \otimes \dots \otimes X_M \otimes \bbbone_{\h_\r} ) = \omega_{\r,M} ( X_1 \otimes \dots \otimes X_M )$,
\item[ii)] (Symmetry) For any bounded operators $X_1,\ldots,X_M$ on $\h_\r$ and any permutation $\pi$ of $\{1,\ldots,M\}$, we have $\omega_{\r,M} (X_1 \otimes \dots \otimes X_M ) = \omega_{\r,M} (X_{\pi(1)} \otimes \dots \otimes X_{\pi(M)})$.
\end{itemize}
A (sequence of) states satisfying these properties obey the hypotheses of the Quantum De Finetti theorem appearing as Theorem 2.1 of \cite{RoNaLe1}:
\medskip

Suppose $\omega_{\r,M}$ satisfies the consistency and symmetry properties above. Then there is a unique Borel probability measure $\mu$ on the set of states $\mathcal S_\r= \{ \omega \in \mathcal B(\h_\r)^{*} \ :\  \omega(\bbbone) = 1 , \,\ \omega \geq 0 \}$ (the set of normalized, positive linear functionals on $\h_\r$), such that for all $M$,
\begin{equation}
\label{definetti}
\omega_{\r,M} = \int\limits_{\mathcal S_\r} d \mu (\omega) \, \omega^{\otimes M},
\end{equation}
where $\omega^{\otimes M}=\omega\otimes\cdots\otimes\omega$ is the $M$-fold product state acting on bounded operators of $\h_\r^{\otimes M}$. 
\medskip

The key property here is that for De Finetti type states, instead of \eqref{20.1} we have 
\begin{equation}
\label{defi1.1}
\lim_{M\rightarrow\infty} \omega_{\r,M}\big(\overline X_{M,1}\cdots \overline X_{M,n}\big) = \int_{\mathcal S_\r} d\mu(\omega)\,  \omega(x_1)\cdots \omega(x_n).
\end{equation}
By linearity of the integral, this allows us to establish the following result.

\begin{prop}
\label{prop:definettidyn}
Suppose $\omega_{\r,M}$ satisfies the consistency and symmetry conditions above, so that the De Finetti theorem applies and gives \eqref{definetti}. Suppose that $G$ and $v$ are bounded operators. Then the reduced system dynamics is given by
\begin{equation}
\label{definetti3}
\rho_\s(t) = \int_{\s_\r} d\mu(\omega) U(\omega,t) \rho_\s(0) U^*(\omega,t),
\end{equation}
where for all $\omega\in\mathcal S_\r$,
$$
i\partial_t U(\omega,t) = H_{\s,\rm eff}(\omega,t) U(\omega,t)\qquad \mbox{with}\qquad 
H_{\s,\rm eff}(\omega,t) = H_\s +\omega(v(t)) G.
$$
\end{prop}

We give a proof of Proposition \ref{prop:definettidyn} in Section \ref{sec:definetti}. It is very similar to the proof in the case the reservoir is in a product state, leading to the result of Theorem \ref{thm1}. However, in contrast to that result, here the effective system dynamics is no longer given by a unitary group. Indeed, by taking the integral in \eqref{definetti3}, the product property of the propagators is broken and the evolution is no longer necessarily Markovian. An analogous situation is described in \cite{CFFM}. There the dynamics of an open system is derived to be given by \eqref{definetti3} with a measure emerging from the classical limit of the reservoir. In that reference, decoherence and non-Markovianity of the system dynamics are investigated.

\section{Applications}
\label{sec:appli}

In Section \ref{sec:res} we describe some reservoir models which fit our assumptions. Then we present a discussion on the effective system dynamics in Section \ref{subs:Stark}.

\subsection{Examples of reservoirs}
\label{sec:res}

We give some examples of reservoir models which satisfy the condition \eqref{10}.
\medskip

{\bf 1. The reservoir as a spin lattice.}  The index $m=1,\ldots M$ labels the position on a lattice (in any dimension) and $M\rightarrow\infty$ corresponds to the infinite size lattice limit (see also Figure \ref{figure1} depicting a two-dimensional lattice). At each lattice site $m$ is located a spin (generally, any quantum system with Hilbert space $\h_\r$, $\dim\h_\r=d_\r<\infty$). The state $\omega_\r$ can be taken as an arbitrary state of a $d_\r$-level system, determined by an arbitrary density matrix $\rho_\r$ on $\h_\r$ and the Hamiltonian $H_\r$ is any hermitian operator on $\h_\r$. $\omega_{\r,M}$ can then be taken as the $M$-fold product of $\omega_\r$, or it may contain correlations, such as in \eqref{stateRM}. Let $X$ be an operator on $\h_\r\otimes\cdots\otimes\h_\r$ ($M$-fold product). As $\omega_{\r,M}$, \eqref{stateRM} is a state, that is, a linear functional of norm one, we have 
\begin{equation}
|\omega_{\r,M}(X)|\le \|X\| 
\end{equation}
where $\|X\|$ is the operator norm of $X$. In view of \eqref{10} we take $X=\oV_{\r,M}(t_1)\cdots \oV_{\r,M}(t_n)$, so
$$
|\omega_{\r,M}(X)|\le \frac{1}{M^n}\sum_{m_1=1}^M\cdots \sum_{m_n=1}^M \|v(t_1)^{[m_1]}\|\cdots \|v(t_n)^{[m_n]}\| = \|v\|^n,
$$
where we used that $\|v(t)^{[m]}\|= \|v\|$. Consequently \eqref{10} holds with $C_{\r,n}=\|v\|^n$ and $b_\r(t)=1$.

What kind of time dependent potentials in the sysetem Hamiltonian can we get through the mean-field coupling for $d_\r=\dim \h_\r<\infty$? Denote the (single site) reservoir Hamiltonian by $h_{\r} = \sum_{l=1}^{d_\r} \lambda_l |e_l\rangle \langle e_l|$ and denote matrix elements of a single site operator $X$ as $[X]_{k,l}=\langle e_k|X|e_l\rangle$. Then
$$
\omega_\r\big(v(t)\big) = \tr_\r \big( \rho_\r v(t) \big) = \sum_{l,k=1}^{d_\r} [\rho_\r]_{kl} [v]_{lk} e^{it(\lambda_l - \lambda_k)},
$$ 
which gives a quasi-periodic time dependent potential in the effective system Hamiltonian $H_{\rm eff}(t)$, \eqref{18.1}.
\medskip
    
{\bf 2. The reservoir as an oscillator field.} Consider $\h_\r$ to be the Hilbert space of a quantum oscillator with creation and annihilation operators $a^\dag, a$, satisfying the commutation relation $[a,a^\dag]=\bbbone$. For the single oscillator Hamiltonian ({\it c.f.} after \eqref{4}) take 
\begin{equation}
\label{hr}
h_\r = \Omega_\r a^\dag a
\end{equation}
where $\Omega_\r>0$. As a first choice, take the interaction operator to be the field operator,
\begin{equation}
\label{v1}
v_1=\varphi = \frac{a^\dag +a}{\sqrt 2}, \quad 
v_1(t) = \varphi(t) = e^{i t h_\r}\varphi e^{-i t h_\r} = \frac{e^{i\Omega_\r t}  a^\dag +e^{-i\Omega_\r t} a}{\sqrt 2}.
\end{equation}
If $\omega_\r$ is chosen to be a gauge-invariant state of the oscillator, meaning that the expectation of a product of an odd number of $a$ or $a^\dag$ vanishes (such as the equilibrium state, for example), then $\omega_\r(v(t))=0$. This means that the effective evolution of $\s$ is not affected by the mean field coupling to the reservoir. We discuss two other models, in which the effective dynamics does feel the coupling. 

\begin{itemize}
\item[\bf (2a)] Coherent state: Suppose $\rho_{\r,M}=|\alpha\rangle\langle\alpha|\otimes\cdots \otimes|\alpha\rangle\langle\alpha|$ is the product of $M$ coherent states $|\alpha\rangle$, $\alpha \in\mathbb C$ defined by $a|\alpha\rangle =\alpha|\alpha\rangle$. Let $v_1$, \eqref{v1} be the interaction operator.

\begin{prop}
\label{prop:2a}
The bound \eqref{10} holds with 
\begin{equation}
\label{Cncoherent}
C_{\r,n} =  (2 \lfloor n/2 \rfloor)!!\, \big(\frac{1}{2}+|\alpha|\big)^n,\qquad b_\r(t)=1,
\end{equation}
where for any even integer $k$, we set $k!!=(k-1) (k-3)\cdots 3\cdot 1$.

\end{prop}

 Note that $2 \lfloor n/2 \rfloor=n$ if $n$ is even and $2 \lfloor n/2 \rfloor=n-1$ if $n$ is odd. By using the root test for the convergence of the series, one shows that \eqref{cond} is satisfied if for some $\xi \ge 0$ we have $C_{\s,n}\le e^{\xi n}$ (which is the case if $\|G\|<\infty$ in particular). We prove Proposition \ref{prop:2a} in Section \ref{sect:prop2a}.

\item[\bf (2b)] Instead of the $v_1=\varphi$ considered in the previous point, now consider the interaction operator 
$$
v_2  = \nu a^\dag a,
$$
where $\nu\in\mathbb R$. This represents a `scattering' interaction in which field quanta are not created or destroyed, as $v_2$ commutes with the number operator $a^\dag a$. 

\begin{prop}
\label{prop:2b}
The bound \eqref{10} is satisfied provided $\omega_{\r,M}$ is product state $\omega_{\r,M}=\omega_\r\otimes\cdots\otimes\omega_\r$ satisfying 
\begin{equation}
\label{44.1}
\omega_\r\big( (a^\dag a)^k\big)\le c^k,\qquad k\in\mathbb N,
\end{equation}
for some constant $c$. In \eqref{10} we can take $C_{\r,n} = (cv)^n$ and $b_\r(t)=1$.
\end{prop}
The condition \eqref{44.1} holds in particular if the density matrix of $\omega_\r$ is a mixture of states all of which have at most $c$ excitations. We prove Propostion \ref{prop:2b} in Section \ref{sect:prop2b}.

The operator $v_2$ is stationary under the Heisenberg dynamics generated by $h_\r$, \eqref{hr}. The effective Hamiltonian for the system dynamics becomes (see \eqref{18}),
\begin{equation}
H_{\rm eff} = H_\s + \nu\omega_\r(a^\dag a)G.
\end{equation}
\end{itemize}

{\bf 3. The reservoir as a quantum field.} Instead of having a single oscillator at each `site' $m$ we may have a collection of them. We now consider $\h_\r=\mathcal F(L^2(\mathbb R^d,d^dk))$, the (bosonic) Fock space over $L^2(\mathbb R^d,d^dk)$. It carries creation and annihilation operators $a^\dag_k, a_k$ and smoothed out versions,
$$
a^\dag(f) = \int_{\mathbb R^d} f(k) a^\dag_k \, d^dk,\quad a(f) = \int_{\mathbb R^d} \overline{f(k)} a^\dag_k \,d^dk,
$$
where $f\in L^2(\mathbb R^d,d^dk)$ is a `test function'. The canonical commutation relations are $[a(f),a^\dag(g)] = \langle f,g\rangle$, where the right side is the inner product in $L^2(\mathbb R^d,d^dx)$. The field Hamiltonian is given by the Hamiltonian
\begin{equation}
h_\r = d\Gamma(\omega) \equiv \int_{\mathbb R^d} \omega(k) a^\dag_k  a_k\, d^dk,
\end{equation}
where $\omega(k)\ge 0$ is the `dispersion relation' of the field. It generates the Bogoliubov tranformation,
$$
e^{it h_\r} a^\dag(f) e^{-it h_\r} = a^\dag(e^{i\omega t}f).
$$
As in the previous oscillator case, the mean field interaction with an operator 
\begin{equation}
\label{vcohfield}
v_1=\varphi(g)\equiv \frac{1}{\sqrt 2}(a^\dag(g)+a(g))
\end{equation}
which is linear in creation and annihilation operators, does not affect the system dynamics if the reservoir state $\omega_\r$ is a gauge-invariant state (such as an equilibrium state), because for those states we have $\omega_\r( e^{i t h_\r}\varphi(g)e^{-i t h_\r})=0$. We then consider again the cases (a) and (b) as in the single oscillator model.

\begin{itemize}
\item[\bf (3a)] Coherent state: Suppose $\rho_{\r,M} =|\Psi_f\rangle\langle\Psi_f|\otimes\cdots\otimes |\Psi_f\rangle\langle\Psi_f|$ is the product of $M$ coherent states $\Psi_f = W(f)|\Omega\rangle$, where $|\Omega\rangle$ is the vacuum state,  $f\in L^2(\mathbb R^d, d^dk)$ is a fixed function and $W(f) = e^{i\varphi(f)}$ is the Weyl operator. We take the interaction operator $v_1$, \eqref{vcohfield}.  As in the oscillator case, we obtain the following result:

\begin{prop}
    \label{prop:3a}
The bound \eqref{10} holds with $C_{\r,n} = (2 \lfloor n/2 \rfloor)!! (\frac{1}{2}+\|f\|)^n \|g\|^n $ and $b_\r(t)=1$.
\end{prop}

A proof of Proposition \ref{prop:3a} can be obtained by following exactly the lines of the proof of Proposition \ref{prop:2a}, see at the end of Section \ref{sect:prop2a}. 

\item[\bf (3b)] In view of a `scattering' interaction, consider the field interaction operator to be given by
$$
v_2 =  a^\dag(g)a(g),\quad v_2(t) =  a^\dag(e^{i\omega t}g) a(e^{i\omega t}g)
$$
for a fixed $g\in L^2(\mathbb R^d, d^dk)$.

\begin{prop}
\label{prop:3b}
The bound \eqref{10} holds provided  $\omega_{\r,M}$ is the $M$-fold tensor product of $\omega_\r$ satisfying ({\it c.f.} \eqref{44.1})
\begin{equation}
\label{44.1.1}
\omega_\r\big(\widehat N^k\big)\le c^k,\qquad k\in\mathbb N,
\end{equation}
for some constant $c$, and where $\widehat N = \int_{\mathbb R^d}a^\dag_ka_k d^dk$ is the number operator. For the constants in \eqref{10} we can take $C_{\r,n} = (2c\|g\|^2)^n$ and $b_\r(t)=1$. 
\end{prop}

We give a proof of Proposition \ref{prop:3b} in Section \ref{sect:prop3b}. The effective Hamiltonian of the system is
\begin{equation}
\label{heff50}
H_{\rm eff}(t) = H_\s + \omega_\r\big(a^\dag(e^{i\omega t}f)a(e^{i\omega t}f)\big) G.
\end{equation}
As an example, consider $\omega_\r = |a^\dag(h)\Omega\rangle\langle a^\dag(h)\Omega|$ for a function $h\in L^2(\mathbb R^d, d^dk)$, where $|\Omega\rangle$ is the vacuum state. This is a state having one excitation (in the form of the wave function $h$). Then
\begin{equation}
\label{51.1}
\omega_\r\big(a^\dag(e^{i\omega t}f)a(e^{i\omega t}f)\big) = \big|\langle h, e^{i\omega t}f\rangle\big|^2.
\end{equation}
Typically, the quantity \eqref{51.1} tends to zero for large times: Take for instance $d=3$, $\omega(k)=|k|$, then (in spherical coordinates $(\Sigma,r=|k|)$ of $\mathbb R^3$),
\begin{equation}
\langle h, e^{i\omega t}f\rangle = \int_0^\infty r^2 e^{irt} \overline{h'(r)} f'(r) dr, 
\end{equation}
where $f'(r) = \int_{S^2} f(r,\Sigma)d\Sigma$ (angles integrated out). By the Riemann-Lebesgue theorem we have $\lim_{t\rightarrow\infty} \langle h,e^{i\omega t}f\rangle=0$ at a speed which depends on the regularity and the infra-red behaviour (small $|k|$) of the functions $f,h$. Compared the the case of a single oscillator in the previous paragraph (or a finite amount of oscillators), the continuous modes of the oscillator field here generates decay in the reservoir dynamics. As a consequence the Hamiltonian $H_{\rm eff}(t)$, \eqref{heff50} converges to the free Hamiltonian $H_\s$ as $t\rightarrow\infty$ (while it is a quasi-periodic function of time for finitely many oscillators). 
\end{itemize}

\subsection{Examples of the effective system dynamics}
\label{subs:Stark}

If $\omega_\r$ is stationary with respect to the free reservoir dynamics then the effective system Hamiltonian $H_{\s,\rm eff} = H_\s+\omega_\r(v) G$ is time-independent, and we  have
$$
\rho_\s(t) =e^{-i t (H_\s +\omega_\r(v) G)}\rho_\s(0) e^{i t (H_\s +\omega_\r(v) G)},
$$
yielding a unitary dynamics with renormalized energies. If the average of the operator $v$ in the state $\omega_\r$ vanishes, then the system dynamics is not affected by the coupling to the reservoir. For finite-dimensional systems $\s$ in contact with a stationary reservoir, the coupling to the reservoir thus simply causes a shift of the energy eigenvalues. If the term $\omega_\r(v) G$ is small relative to $H_\s$, then the uncoupled energies are modified little by the interaction with the reservoir. However, our result is not of perturbative nature, in the sense that we do not make any assumptions on the relative sizes of $H_\s$ and $\omega_\r(v) G$. The interaction term $\omega_\r(v) G$ may be of the same size as $H_\s$, or it may exceed it arbitrarily.

In case $\dim\h_\s<\infty$, the spectrum of $H_{\s,\rm eff}(t)$ consists only of eigenvalues (`bound states'). However, in the infinite dimensional case, the contact with the reservoir can modify the nature of the spectrum: it can turn bound states into scattering states and {\it vice versa}. In other words, the effective potential which $\s$ is subjected to can trap or `localize' particles or on the contrary, it can `ionize' trapped particles. This is what we explore now.

One of the prime examples of continuous systems is that of $N=1,2,\ldots$ quantum particles in position space $\mathbb R^d$ ($d=1,2,3,\ldots$). The Hilbert space of a single particle is $\h_\s=L^2(\mathbb R^d,d^dx)$ and the free Hamiltonian is the Laplacian, $H_\s=-\Delta=-\sum_{j=1}^d \partial^2_{x_j}$. Consider an interaction operator $G=V(x)$ to be a bounded, real valued function on $\mathbb R^d$ ($G$ is a multiplication operator). Assuming the reservoir to be in a stationary state, Theorem \ref{thm1} shows that the effective Hamiltonian of the particle is given by
\begin{equation}
H_{\rm eff} = -\Delta +W(x),
\end{equation}
which is a Schr\"odinger operator with potential $W(x)=\omega_\r (v) V(x)$. Let $W(x)$ be of the form of a potential well. Then it is known that $H_{\rm eff}$ generally has (depending on the depth of the well an arbitrary number of) {\it bound states}, while the Laplacian alone, of course, does not have any eigenvalues. This shows that the mean field coupling to the reservoir can create bound states, that is, it can cause localization in the particle system. Conversely, one could start with a system Hamiltonian $H_\s=-\Delta +W(x)$ which does have bound states -- then the mean field coupling will alter the potential and can create additional bound states or remove some or all of them.  

In the above examples we have considered bounded potentials, $\|G\|=\sup_{x\in\mathbb R^d}|V(x)|<\infty$, so that the condition \eqref{10.2} is easily seen to be correct (see also the explanation after \eqref{cond}). For such potentials, the interaction with the reservoir can change the number of bound states only by a finite amount. We discuss now an interaction with the field which produces an unbounded potential.

Consider a particle in $\mathbb R^d$, with $\h_\s=L^2(\mathbb R^d, d^dx)$ with free Hamiltonian $H_\s =-\Delta$ and an interaction operator 
\begin{equation}
\label{61}
G= \hat E\cdot x,
\end{equation}
the operator of multiplication by $\hat E\cdot x$ (unbounded potential), where $\hat E\in \mathbb R^d$ is a fixed vector of unit length and $x$ is the operator of multiplication by $x\in\mathbb R^d$. We assume that the reservoir state $\omega_\s$ and interaction operator $v$ satisfy $\omega_\r(v(t))=\alpha\in\mathbb R$ (independent of time) but do not further specify the reservoir characteristics. The resulting effective system Hamiltonian is
\begin{equation}
H_{\rm eff} = -\Delta + \alpha \hat E\cdot x,
\end{equation}
which is the `Stark Hamiltonian' describing a free charged particle in a constant electric field $\hat E$. 

\begin{prop}
\label{prop:Stark}
Suppose $G$ is given in \eqref{61} and that $\rho_\s = \sum_{\alpha\ge 0}p_\alpha |\psi_\alpha\rangle\langle\psi_\alpha|$ is a density matrix on $\h_\s =L^2(\mathbb R^d, d^dx)$, with probabilities $p_\alpha$ and normalized vectors $\psi_\alpha$. Suppose that there is a constant $c$ such that for all $k\in\mathbb N$, 
\begin{equation}
\label{regularity}
\big\|\big(\hat E\cdot (x-\nabla_x) \ \hat E\cdot (x+\nabla_x) \big)^k \psi_\alpha\big\|\le c^k.
\end{equation}
Take a reservoir satisfying \eqref{10} with $C_{\r,n}\le (c')^n$ for some constant $c'$ and $b_\r(t)$ a locally integrable function of $t\ge 0$. Then the system dynamics is given by the effective Hamiltonian
\begin{equation}
H_{\rm eff}(t)  =-\Delta +\omega_\r(v(t))\, \hat E\cdot x.
\label{72.1}
\end{equation}
\end{prop}

In dimension $d=1$, the condition \eqref{regularity} is $\|(-\Delta +x^2-1)^k\psi_\alpha\|\le c^k$. The eigenfunctions of $-\Delta+x^2-1$ are the Hermite functions $\Psi_\nu$, $\nu=0,1,2,\ldots$, satisfying $(-\Delta+x^2-1)\Psi_\nu = 2\nu\Psi_\nu$. They are an orthonormal basis of $L^2(\mathbb R, dx)$. Let $\nu_*$ be any fixed integer and for each $\alpha$, take $\psi_\alpha$ to be a linear combination of $\Psi_\nu$ with $\nu\le \nu_*$. Then \eqref{regularity} holds with $c=\nu_*$. One may carry out a similar analysis in dimension $d\ge 2$. The condition \eqref{regularity} can be given directly on the density matrix by requiring that 
$$
\tr_{\mathcal{H}_\s} \big( \rho_\s \big(\hat E\cdot (x-\nabla_x) \ \hat E\cdot (x+\nabla_x) \big)^{2k} \big) \leq c^{2k},\quad k\in\mathbb N.
$$

For each $t$ fixed, $H_{\rm eff}(t)$,  \eqref{72.1} has absolutely continuous spectrum and no eigenvalues (the electric field drives the charged particle indefinitely in one direction). One may consider the problem on a half line, $\h_\s = L^2({\mathbb R}_+,dx)$ with Hamiltonian $H_\s=-\Delta$ with Dirichlet boundary conditions \cite{Tolosa-Uribe-2022}. The result of Proposition \ref{prop:Stark} is still valid  (our proof directly carries through to this case). It is well known that the spectrum of the resulting effective Hamiltonian \eqref{72.1}, for $\omega_\r(v(t)) E>0$, is purely discrete \cite{Reed-Simon-IV} (Theorem XIII.16). As a consequence, when the free particle on the half-line is coupled to the reservoir,  the nature of its dynamics changes entirely, from having only scattering states to having only bound states.

\section{Extensions of Theorem \ref{thm1} }
\label{sec:extensions}

In this section we present some generalizations of Theorem \ref{thm1}.
\medskip

{\bf  1. Many-site (many-body) system-reservoir interaction.} Consider a $\nu$-body reservoir operator $v^\nu$, that is an  operator acting on the $\nu$-fold tensor product $\h_\r\otimes\cdots\otimes\h_\r$ and let the interaction operator $V_M$ in \eqref{3} be given by
\begin{equation}
V_M = G\otimes  {M\choose \nu}^{-1} \sum_{\Lambda\in \mathcal P_{\nu,M}} (v^\nu)^{[\Lambda]},
\label{19.1}
\end{equation}
where $\Lambda$ is drawn from the set  $\mathcal P_{\nu,M}$ of all clusters of $\nu$ sites (indices) of the $M$ total reservoir sites, and where $(v^\nu)^{[\Lambda]}$ operates on the  cluster $\Lambda$. In \eqref{19.1}, we have replaced the normalization factor $M^{-1}$ in \eqref{M}  by ${M\choose \nu}^{-1}$ which divides by the total number of ways to choose clusters of size $\nu$ among $M$ elements. Suppose that $\omega_{\r,M}$ is of product form $\omega_{\r,M}=\omega_\r\otimes\cdots\otimes\omega_\r$. Then \eqref{12} is valid with the unitary solving 
\begin{equation}
\label{20.2}
i\partial_t U(t) = \Big(H_\s +\omega^{\otimes\nu}_\r\big(v^\nu(t)\big)\, G\Big) U(t),\quad U(0)=\bbbone,
\end{equation}
where $\omega^{\otimes\nu}_\r = \omega_\r\otimes \cdots\otimes\omega_\r$ ($\nu$ factors). 
We give the proof of this result within the proof of Theorem \ref{thm1}. 
\medskip

{\bf 2. Interaction as a sum of terms.} Take an interaction operator given by a sum of terms of the form \eqref{M}, 
\begin{equation}
V_M = \sum_{j=1}^J G_j\otimes \oV_{\r,M,j},\qquad \oV_{\r,M,j} =  \frac1M \sum_{m=1}^M (v_j)^{[m]},
\end{equation}
where $G_j$ and $v_j$ are operators  such that \eqref{10}, \eqref{cond} are satisfied. Then \eqref{12} is valid with the unitary being the solution of  
\begin{equation}
i\partial_t U(t) = \Big(H_\s + \sum_{j=1}^N \omega_\r\big(v_j(t)\big) G_j \Big) U(t),\quad U(0)=\bbbone.	
\label{24.2}
\end{equation}
A proof of this result is obtained along the lines of the proof of Theorem \ref{thm2} below, where we show that the result \eqref{24.2} is also valid when $\s$ is composed of $N$ subsystems and the operators $G_j$ are given by $G_j \equiv \bbbone_{\h_{\s,1}}\otimes\cdots\otimes G_j\otimes \cdots\otimes \bbbone_{\h_{\s,N}}$, acting on the $j$th subsystem of $\s$. 
\medskip

{\bf  3. Macroscopically varying reservoir state.} Suppose the state $\rho_{\r,M}$ is a product state in which $M_k$ factors are in given states $\sigma_{\r,k}$ on $\h_\r$, for $k=1,\ldots,K$. We have $M_1+\cdots+M_K=M$ (so the $M_k$  depend on $M$) and we set
$$
\lambda_k = \lim_{M\rightarrow\infty}\frac{M_k}{M} \in [0,1].
$$
As $\sum_{k=1}^K\lambda_k=1$ the $\lambda_k$ are probabilities and we set
\begin{equation}
\label{mi23}
\omega_\r(\cdot ) = {\rm Tr}(\rho_\r \cdot ),\qquad 
\rho_\r = \sum_{k=1}^K \lambda_k \,\sigma_{\r,k}.
\end{equation}
Then Theorem \ref{thm1} holds with $\omega_\r$ given in \eqref{mi23}.

To show this, we proceed as in the proof of Theorem \ref{thm1} with the following modification. The condition \eqref{corrstate} does not hold for the state described here:  Even though $\rho_{\r,M}$ is a product state the  factors are not the same. We thus need an alternative proof of the result \eqref{20.1} of Proposition \ref{prop1'}. Namely, we show now that (using the notation of Proposition \ref{prop1'}) 
\begin{equation}
\label{mi24}
    \lim_{M \rightarrow \infty } \tr_{\r,M} \big( \rho_{\r,M} \overline X_{M,1} \cdots \overline X_{M,n} \big) = \omega_\r(x_1)\cdots \omega_\r(x_n),
\end{equation}
where $\omega_\r$ is given in \eqref{mi23}. The proof of Theorem \ref{thm1} is then the same as before.

To show \eqref{mi24} we write the left side as
\begin{equation}
\lim_{M\rightarrow\infty} \frac{1}{M^n} \sum_{m_1,\dots, m_n=1}^M \tr_{\r,M} \big( \rho_{\r,M} x_1^{[m_1]} \cdots x_n^{[m_n]} \big)  = \lim_{M\rightarrow\infty} \frac{1}{M^n} \sum_{\mathcal M_n} \tr_{\r,M} \big( \rho_{\r,M} x_1^{[m_1]} \cdots x_n^{[m_n]} \big),
\label{mi25}
\end{equation}
where $\mathcal M_n$ is the set of all distinct indices, as in \eqref{21.1}. For each $( m_1, \dots, m_n ) \in \mathcal{M}_n$, there are $l_1$ of the indices which `point to' a site where the state $\rho_{\r,M}$ has the factor $\sigma_{\r,1}$. There are $l_2$ indices corresponding to $\sigma_{\r,2}$, and so on. This defines numbers $0\le l_k\le n$, $k=1,\ldots,K$, satisfying $l_1+\cdots +l_K=n$. For $l_1,\ldots,l_K$ fixed, the expectation value (trace in \eqref{mi25}) is given by 
$$
\sum_{\pi \in \mathcal{P}_{l_1,\dots, l_K}} \prod_{q=1}^n \tr_{\r} \big( \sigma_{\r,\pi(q)} \, x_{q} \big),
$$
where 
\begin{equation}
\label{Pl}
\mathcal{P}_{l_1,\dots, l_K} = \big\{ \pi : \{1, \dots, n \} \rightarrow \{ 1, \dots , K \} \mbox{\  such that $|\pi^{-1}(k)|=l_k$ for $k=1,\ldots,K$} \big\}.
\end{equation}
Additionally, when distributing the $m_1,\ldots,m_n$ into the groups defined by $l_1,\ldots,l_K$, their `location' within the group creates a combinatorial factor. For the first group this factor is $M_1(M_1-1)\cdots (M_1-l_1+1)=\frac{(M_1)!}{(M_1-l_1)!}$, and similarly for the other groups. We thus obtain
\begin{equation*}
    \frac{1}{M^n}  \sum_{\mathcal M_n}  \tr_{\r,M} \big( \rho_{\r,M} x_1^{[m_1]} \cdots x_n^{[m_n]} \big)  
     = \sum_{\substack{l_1,\dots,l_K \geq 0 \\  l_1+\cdots +l_K=n }} \frac{\prod_{k=1}^K \frac{(M_k)!}{(M_k - l_k)!}}{M^n}   \sum_{\pi \in \mathcal{P}_{l_1,\dots, l_K}} \prod_{q=1}^n \tr_{\r} \big( \sigma_{\r,\pi(q)} x_{q} \big).
\end{equation*}
Next, $\lim_{M\rightarrow\infty} M^{-l_k}\frac{(M_k)!}{(M_k-l_k)!}=\lambda_k^{l_k}$ and therefore,
\begin{align*}
\lim_{M\rightarrow\infty} \frac{1}{M^n} \sum_{\mathcal M_n}  \tr_{\r,M} \big( \rho_{\r,M} x_1^{[m_1]} \cdots x_n^{[m_n]} \big) 
    = \sum_{\substack{l_1,\dots,l_K \geq 0  \\ l_1+\cdots +l_K=n }} \lambda_1^{l_1}\cdots\lambda_K^{l_K} \sum_{\pi \in \mathcal{P}_{l_1,\dots, l_K}} \prod_{q=1}^n \tr_{\r} \big( \sigma_{\r,\pi(q)} x_{q} \big).
\end{align*}
The double sum over the decomposition $l_1+\cdots+l_K=n$ and then over the functions $\pi\in \mathcal P_{l_1,\ldots,l_K}$ constraint to have the sizes of the inverse images as dictated by \eqref{Pl} is the same as the sum over all functions $f:\{1,\ldots,n\}\rightarrow \{1,\ldots,K\}$. Thus, distributing the product of the $\lambda_k$ over the states $\sigma_{\r,k}$, we get 
\begin{align}
\label{mi28}
    \sum_{\substack{l_1,\dots,l_K \geq 0  \\ l_1+\cdots +l_K=n }} & \lambda_1^{l_1}\cdots\lambda_K^{l_K} \sum_{\pi \in \mathcal{P}_{l_1,\dots, l_K}} \prod_{q=1}^n \tr_{\r} \big( \sigma_{\r,\pi(q)} x_{q} \big) =  \sum_{f : \{1, \dots, n\} \rightarrow \{1, \dots, K\} } \prod_{q=1}^n \tr_{\r} \big( \lambda_{f(q)} \sigma_{\r,f(q)} x_{q} \big). 
\end{align}
Finally, we note that the right side of \eqref{mi28} is just $\prod_{q=1}^n \big( \sum_{j=1}^K \lambda_j \tr_{\r} \big( \sigma_{\r,j} x_q \big) \big)$.  Here is a way of seing this: Set $ a_{q,j} : = \lambda_j \tr_{\r} \big( \sigma_{\r,j} x_q \big)$ and write down the product
\begin{equation*}
    (a_{1,1} + \dots + a_{1,K } ) \cdot (a_{2,1} + \dots + a_{2,K } ) \cdots (a_{n,1} + \dots + a_{n,K } ).
\end{equation*}
By multiplying out the terms in the product, we have to pick one of the $a_{q,j}$ from each of the parenthesis indexed by $q$. But this is exactly constructing a function $f$: we assign $f(q)=j$ by picking the $j$-th element from the $q$-th parenthesis. We have thus shown that 
\begin{equation}
    \lim_{M \rightarrow  \infty } \tr_{\r,M} \big( \rho_{\r,M} \overline X_{M,1} \cdots \overline X_{M,n} \big) = \prod_{q=1}^n \big( \sum_{j=1}^K \lambda_j \tr_{\r} ( \sigma_{\r,j} x_q ) \big),
\end{equation}
that is, \eqref{mi24} holds with $\omega_\r$ given in \eqref{mi23}.

\section{Proofs}
\label{sec:proofs}

We collect the proofs of our main results, Theorems \ref{thm1} and \ref{thm2} and some Propositions. 

\subsection{Proof of Theorem \ref{thm1}}
We start by using the Dyson series expansion,
\begin{eqnarray}
\lefteqn{
{\rm Tr}_{\r,M} \big( e^{it H_M^0} e^{-it H_{\s\r,M}} \rho_{\s\r,M}\, e^{it H_{\s\r,M}}e^{-it H_M^0}\big)}\nonumber\\
&=& \sum_{n\ge 0}(-i)^n \int_{0\le t_n\le\cdots \le t_1\le t} dt_1\cdots dt_n {\rm Tr}_{\r,M}\big( 
\big[V_M(t_1), [ \cdots [V_M(t_n),\rho_\s\otimes\rho_{\r,M}]\ldots ]\big]\big),
\label{13}
\end{eqnarray}
where we set 
\begin{equation}
\label{13.1}
H_M^0 = H_\s + H_{\r,M}
\end{equation}
and the integrand is the multiple commutator of the initial state with the interaction operator at different times. We first show that the series converges in trace norm of system operators, for all $t$, and uniformly in $M$.  Expanding the multiple commutator yields $2^n$ terms, each of the form
\begin{equation}
\label{16.1}
G(t_{\ell_1})\cdots G({t_{\ell_L}}) \rho_\s G(t_{\ell_L+1})\cdots G(t_{\ell_n}) \otimes \oV_{\r,M}(t_{\ell_1})\cdots \oV_{\r,M}(t_{\ell_L})\rho_{\r,M} \oV_{\r,M}(t_{\ell_L+1})\dots \oV_{\r,M}(t_{\ell_n})
\end{equation}
for some $L$ and some permutation $j\mapsto \ell_j$ of the $n$ indices. Due to the cyclicity of the trace and \eqref{10}, we have 
\begin{equation}
\label{15}
\Big| {\rm Tr}_{\r,M} \big(\oV_{\r,M}(t_{\ell_1})\cdots \oV_{\r,M}(t_{\ell_L})\rho_{\r,M} \oV_{\r,M}(t_{\ell_L+1})\dots \oV(t_{\ell_n})\big)\Big| \le C_{\r,n}\, b_\r(t_1)\cdots b_\r(t_n).
\end{equation}
Further, according to \eqref{10.2}, 
\begin{equation}
\label{16}
\| G(t_{\ell_1})\cdots G({t_{\ell_L}}) \rho_\s G(t_{\ell_L+1})\cdots G(t_{\ell_n}) \|_1\le C_{\s,n}\, b_\s(t_1)\cdots b_\s(t_n)
\end{equation}
Using the bounds \eqref{15} and \eqref{16} we obtain
$$
\big\| {\rm Tr}_{\r,M}\big( 
\big[V_M(t_1), [ \cdots [V_m(t_n),\rho_\s\otimes\rho_{\r,M}]\ldots ]\big]\big)\big\|_1\le C_{\s,n} C_{\r,n} b_\s(t_1)\cdots b_\s(t_n) b_\r(t_1)\cdots b_\r(t_n).
$$
The multiple time integral in \eqref{13} is bounded in the trace norm by
\begin{align}
\Big\|\int_{0\le t_n\le \cdots \le t_1\le t}   dt_1\cdots dt_n {\rm Tr}_{\r,M}\big( 
\big[V_M(t_1), [ \cdots [V_M(t_n),\rho_\s\otimes\rho_{\r,M}]\ldots ]\big]\big)\Big\|_1 \le \nonumber\\
\le 2^n C_{\s,n} C_{\r,n} \int_{0\le t_n\le \cdots \le t_1\le t}   dt_1\cdots dt_n b_\s(t_1)b_\r(t) \cdots b_\s(t_n)b_\r(t_n) = \nonumber\\
= \frac{2^n C_{\s,n} C_{\r,n}}{n!} \Big[\int_0^t b_\s(s)b_\r(s)ds \Big]^n = \frac{C_{\s,n} C_{\r,n}\, (2 B(t))^n}{n!},
\label{38}
\end{align}
where we used the symmetry of the integrand  to tranform the integral over the simplex into an integral over a hypercube. The factor $2^n$ comes from expanding the multiple commutator into $2^n$ terms. The bound \eqref{38} together with \eqref{cond} shows that the series \eqref{13} converges for any value of  and $t$, and uniformly in $M$. As the series (and the integral over times) converges uniformly in $M$, we have 
\begin{multline}
\lim_{M\rightarrow\infty} 
{\rm Tr}_{\r,M} \big( e^{it H_M^0} e^{-it H_{\s\r,M}} \rho_{\s\r,M}\, e^{it H_{\s\r,M}}e^{-it H_M^0}\big)\\
= \sum_{n\ge 0}(-i)^n \int_{0\le t_n\le\cdots \le t_1\le t} dt_1\cdots dt_n \lim_{M\rightarrow\infty} {\rm Tr}_{\r,M}\big( 
\big[V_M(t_1), [ \cdots [V_M(t_n),\rho_\s\otimes\rho_{\r,M}]\ldots ]\big]\big).
\label{19}
\end{multline}
We next simplify the value of the limit in \eqref{19} using Proposition \ref{prop1'}, an argument which is central to many arguments in mean field theory. Namely, we use the result \eqref{20.1} in the expression \eqref{19}, where the role of the $\overline X_j$ is played by the factors $\oV_{\r,M}(t_j)$ in the operators $V_M(t_j) = G(t_j)\otimes \oV_{\r,M}(t_j)$. By expanding out the multi-commutators, as explained above, we obtain $2^n$ terms of the form \eqref{16.1}. Upon taking the partial trace ${\rm Tr}_{\r,M}$ and sending $M\rightarrow\infty$, the term \eqref{16.1} becomes
\begin{multline}
\lim_{M\rightarrow\infty} {\rm Tr}_{\r,M} \Big( G(t_{\ell_1})\cdots G({t_{\ell_L}}) \rho_\s G(t_{\ell_L+1})\cdots G(t_{\ell_n})\\
 \otimes \oV_{\r,M}(t_{\ell_1})\cdots \oV_{\r,M}(t_{\ell_L})\rho_{\r,M} \oV_{\r,M}(t_{\ell_L+1})\dots \oV(t_{\ell_n})\Big)\\
=  G(t_{\ell_1})\cdots G({t_{\ell_L}}) \rho_\s G(t_{\ell_L+1})\cdots G(t_{\ell_n})\omega_\r(v(t_1))\cdots\omega_\r(v(t_n)) \\
= G'(t_{\ell_1})\cdots G'({t_{\ell_L}}) \rho_\s G'(t_{\ell_L+1})\cdots G'(t_{\ell_n}),
\label{44}
\end{multline}
where we set
\begin{equation}
\label{25}
G'(t) = \omega_\r(v(t)) G(t).
\end{equation}
This shows that by taking $\lim_{M\rightarrow\infty} {\rm Tr}_{\r,M}$ in the Dyson series \eqref{19} we obtain the following Dyson series involving the system alone,
\begin{multline}
\lim_{M\rightarrow\infty} 
{\rm Tr}_{\r,M} \big( e^{it H_M^0} e^{-it H_{\s\r,M}} \rho_{\s\r,M}\, e^{it H_{\s\r,M}}e^{-it H_M^0}\big)\\
= \sum_{n\ge 0}(-i)^n \int_{0\le t_n\le\cdots \le t_1\le t} dt_1\cdots dt_n 
\big[G'(t_1), [ \cdots [G'(t_n),\rho_\s]\ldots ]\big].
\label{26}
\end{multline}
The left side of \eqref{26} equals (see 
\eqref{7}, \eqref{13.1})
\begin{equation}
\label{27}
\lim_{M\rightarrow\infty} 
{\rm Tr}_{\r,M} \big( e^{it H_M^0} e^{-it H_{\s\r,M}} \rho_{\s\r,M}\, e^{it H_{\s\r,M}}e^{-it H_M^0}\big)=e^{i t H_
\s} \rho_\s(t) e^{-i t H_\s}.
\end{equation}
We take the derivative with respect to $t$ of both sides in \eqref{26} to obtain
\begin{multline}
e^{i t H_\s}\big( i[H_\s,\rho_\s(t)] +\partial_t\rho_\s(t) \big) e^{- it H_\s}\\
=\sum_{n\ge 1}(-i)^n \int_{0\le t_n\le\cdots \le t_2\le t} dt_2\cdots dt_n 
\big[G'(t), [G'(t_2), [\cdots [G'(t_n),\rho_\s]\ldots ] ]\big]\\
=-i [G'(t), e^{itH_\s}\rho_\s(t)e^{-it H_\s}].
\label{28}
\end{multline}
We used the formulas \eqref{26} and \eqref{27} to arrive at the last equality. It follows from \eqref{28} that 
\begin{equation}
i\partial_t\rho_\s(t) = [H_\s,\rho_\s(t)] + [e^{-it H_\s}G'(t)e^{i t H_\s},\rho_\s(t)].
\end{equation}
From \eqref{25}, \eqref{8} we have $e^{-it H_\s}G'(t)e^{i t H_\s}= \omega_\r(v(t))G$, so
\begin{equation}
i\partial_t\rho_\s(t) = [H_\s+\omega_\r(v(t)) G ,\rho_\s(t)].
\end{equation}
This shows \eqref{12} and completes the proof of Theorem \ref{thm1}. \hfill $\qed$
\bigskip
 
We now give a proof of \eqref{20.2} for the interaction $V_M$ as in \eqref{19.1}. To do so, we repeat the above argument and only need to modify Proposition \ref{prop1'} to the following.

\begin{prop}
\label{prop2}
Suppose $\omega_{\r,M}$ satisfies an analogue condition to \eqref{corrstate} of the form
\begin{equation*}
    \lim_{M \rightarrow \infty } \sup_{\substack{ \Lambda_1 \dots, \Lambda_n \in \mathcal{P}_{\nu,M} \\
    \forall i \neq j , \Lambda_i \cap \Lambda_j = \emptyset }} \big| \omega_{\r,M} \big( (x_1^\nu)^{[\Lambda_1]}   \cdots (x_n^\nu)^{[\Lambda_n]} \big) - \omega_\r^{\otimes \nu}(x_1^\nu) \cdots \omega_\r^{\otimes \nu}(x_n^\nu) \big| = 0
\end{equation*}
and let $x_j^{\nu}$, $j\ge 1$, be operators on $\h_{\r,\nu}$ such that all mixed moments in the state $\omega_{\r,M}$ are finite: For any $n$ there is an $M_0(n)$ and $C(n)$ such that for $M\ge M_0(n)$,
$$
\omega_{\r,M}\big( (x_1^{\nu})^{[\Lambda_1]} \dots (x_n^{\nu})^{[\Lambda_n]} \big) \le C(n).
$$
Define the operators $\overline X_{M,j} = \binom{M}{\nu}^{-1} \sum_{\Lambda \in \mathcal P_{\nu,M}} (x_j^{\nu})^{[\Lambda]}$ on $\h_{\r,M}$. Then we have, for any $n\in\mathbb N$,
\begin{equation*}
\lim_{M\rightarrow\infty} \omega_{\r,M}\big(\overline X_{M,1}\cdots \overline X_{M,n}\big) = \omega_\r^{\otimes \nu} (x_1^{\nu}) \dots \omega_\r^{\otimes \nu} (x_n^{\nu}),
\end{equation*}
where $\omega_\r^{\otimes \nu} (x^{\nu}) = (\omega_\r\otimes\cdots\otimes\omega_\r)(x^\nu) = {\rm Tr}_{\h_{\r,\nu}}\big((\rho_\r\otimes\cdots\otimes\rho_\r) x^{\nu})$. 
\end{prop}

{\it Proof of Proposition \ref{prop2}.\ } We introduce the set $\mathcal M_n$ consisting of all collections $(\Lambda_1,\ldots,\Lambda_n)\in \mathcal (\mathcal P_{\nu,M})^n$ such that $\Lambda_k\cap\Lambda_\ell=\emptyset$ whenever $k\neq \ell$. The cardinality of $\mathcal M_n$ is (for $M\ge n\nu$) 
    \begin{equation*}
        |\mathcal M_n| = \binom{M}{\nu} \binom{M-\nu}{\nu} \dots \binom{M-(n-1)\nu}{\nu}.
    \end{equation*}
It satisfies 
$$ 
\lim_{M \rightarrow \infty} \frac{|\mathcal M_n|}{\binom{M}{\nu}^n} = 1. 
$$ 
Proceeding exactly as in the proof of Proposition \ref{prop1'} the sum 
\begin{equation*}
\frac{1}{\binom{M}{\nu}^n}  \sum\limits_{\Lambda_1, \dots , \Lambda_n \in \mathcal M_n } \omega_{\r,M} \big((x_1^\nu)^{[\Lambda_1]} \dots (x_n^\nu)^{[\Lambda_n]} \big)  
\end{equation*}
converges to $\omega_\r^{\otimes \nu}(x_1^\nu) \cdots \omega_\r^{\otimes \nu}(x_n^\nu)$ in the limit $M\rightarrow\infty$. The remaining part of the sum converges to zero,
$$
\lim_{M\rightarrow\infty} \frac{1}{{M\choose \nu}^n} \sum_{\Lambda_1,\ldots,\Lambda_n\in \mathcal M_n^c} \omega_{\r,M} \big((x_1^\nu)^{[\Lambda_1]} \dots (x_n^\nu)^{[\Lambda_n]} \big)=0,
$$
which is again shown as in Proposition \ref{prop1'}. \hfill $\square$

\subsection{Proof of Theorem \ref{thm2}}

In the Dyson series \eqref{13} we now have
$$
V_M(t) = e^{it (H_\s+H_{\r,M})} V_M e^{-it (H_\s+H_{\r,M})} = \sum_{j=1}^N G_j(t)\otimes \overline V_{\r,M,j}(t),
$$
where $G_j(t) = e^{it H_\s^{(j)}} G_j e^{-it H_\s^{(j)}}$  and $\overline V_{\r,M,j}(t) = e^{it H_{\r,M}} \overline V_{\r,M,j} e^{-it H_{\r,M}}$ (see also \eqref{8}). Therefore,
\begin{eqnarray}
	\lefteqn{
		{\rm Tr}_{\r,M} \big( e^{it H_M^0} e^{-it H_{\s\r,M}} \rho_{\s\r,M}\, e^{it H_{\s\r,M}}e^{-it H_M^0}\big)}\nonumber\\
	&=&  \sum_{n\ge 0}(-i)^n \sum_{j_1=1}^N\cdots \sum_{j_n=1}^N \int_{0\le t_n\le\cdots \le t_1\le t} dt_1\cdots dt_n\nonumber\\
	&&  {\rm Tr}_{\r,M}\big( 
	\big[G_{j_1}(t_1)\otimes \overline V_{\r,M,j_1}(t_1),  [ \cdots [G_{j_n}(t_n)\otimes \overline V_{\r,M,j_n}(t_n),\rho_\s\otimes\rho_{\r,M}]\ldots ]\big]\big).
	\label{13-1}
\end{eqnarray}
For $j_1,\ldots,j_n$ fixed, we estimate in in trace norm (for operators on $\h_\s$), 
\begin{align*}
\Big\|  {\rm Tr}_{\r,M}\big( 
\big[G_{j_1}(t_1)\otimes \overline V_{\r,M,j_1}(t_1),  [ \cdots [G_{j_n}(t_n)\otimes \overline V_{\r,M,j_n}(t_n),\rho_\s\otimes\rho_{\r,M}]\ldots ]\big]\big)\Big\|_1\\
 \le 2^n C_{\s,n} C_{\r,n} \, b_\s(t_1)\cdots b_\s(t_n) b_\r(t_1)\cdots b_\r(t_n),
\end{align*}
where we used the bounds \eqref{10.1} and \eqref{10.3} and the factor $2^n$ comes from expanding the multiple commutator in $2^n$ terms of the form (compare with \eqref{16.1})
\begin{align}
G_{j_1}(t_{\ell_1})\cdots G_{j_L}(t_{\ell_L}) \rho_\s G_{j_{L+1}}(t_{\ell_{L+1}})\cdots G_{j_{L_n}}(t_{\ell_{L_n}}) \nonumber\\
\otimes 
\oV_{\r,M,j_1}(t_{\ell_1})\cdots \oV_{\r,M,j_L}(t_{\ell_L})\rho_{\r,M} \oV_{\r,M,j_{L+1}}(t_{\ell_L+1})\dots \oV_{\r,M,j_n}(t_{\ell_n}).
\label{26.1}
\end{align}
It follows by the same argument given after \eqref{38}, that the series \eqref{13-1} converges uniformly in $M$ due to condition \eqref{cond}. Next, we have, for each $j_1,\ldots, j_n$ fixed,
\begin{align*}
\lim_{M\rightarrow\infty} {\rm Tr}_{\r,M}\Big( \oV_{\r,M,j_1}(t_{\ell_1})\cdots \oV_{\r,M,j_L}(t_{\ell_L})\rho_{\r,M} \oV_{\r,M,j_{L+1}}(t_{\ell_L+1})\dots \oV_{\r,M,j_n}(t_{\ell_n}) \Big)\\
= \omega_\r\big(v_{j_1}(t_{\ell_1})\big) \cdots  \omega_\r\big(v_{j_n}(t_{\ell_n})\big).
\end{align*}
Therefore, in view of \eqref{13-1}, when taking the trace of the reservoir and $M\rightarrow\infty$ and summing over the $j_1,\ldots,j_n$, the term \eqref{26.1} becomes ({\it c.f.} \eqref{44})
\begin{align}
\lim_{M\rightarrow\infty} \sum_{j_1=1}^N\cdots\sum_{j_n=1}^N {\rm Tr}_{\r,M}\Big(G_{j_1}(t_{\ell_1})\cdots G_{j_L}(t_{\ell_L}) \rho_\s G_{j_{L+1}}(t_{\ell_{L+1}})\cdots G_{j_{L_n}}(t_{\ell_{L_n}}) \otimes \nonumber\\
\otimes 
\oV_{\r,M,j_1}(t_{\ell_1})\cdots \oV_{\r,M,j_L}(t_{\ell_L})\rho_{\r,M} \oV_{\r,M,j_{L+1}}(t_{\ell_L+1})\dots \oV_{\r,M,j_n}(t_{\ell_n}) \Big) \nonumber\\
= G''(t_{\ell_1})\cdots G''(t_{\ell_L}) \rho_\s G''(t_{\ell_{L+1}})\cdots G''(t_{\ell_{L_n}}),
\end{align}
where
$$
G''(t) =  \sum_{j=1}^N  \omega_\r\big(v_j(t)\big) G_{j}(t).
$$
As in \eqref{26}, we see that 
\begin{multline}
\rho_\s(t) = \lim_{M\rightarrow\infty} 
	{\rm Tr}_{\r,M} \big( e^{it H_M^0} e^{-it H_{\s\r,M}} \rho_{\s\r,M}\, e^{it H_{\s\r,M}}e^{-it H_M^0}\big)\\
	= \sum_{n\ge 0}(-i)^n \int_{0\le t_n\le\cdots \le t_1\le t} dt_1\cdots dt_n 
	\big[G''(t_1), [ \cdots [G''(t_n),\rho_\s]\ldots ]\big],
	\label{26-1}
\end{multline}
from which \eqref{24.1} follows. This completes the proof of Theorem \ref{thm2}. \hfill $\qed$

\subsection{Proof of Proposition \ref{prop1'}}
\label{sect:prop1'}

We have
\begin{eqnarray}
\omega_{\r,M}\big(\overline X_{M,1}\cdots \overline X_{M,n}\big) &=& \frac{1}{M^n} \sum_{m_1=1}^M\cdots \sum_{m_n=1}^M
\omega_{\r,M}\big(x_1^{[m_1]}\cdots x_n^{[m_n]} \big)\nonumber\\
&=&\frac{1}{M^n} \Big(\sum_{\mathcal M_n} + \sum_{{\mathcal M_n}^c} 
\Big) \omega_{\r,M}\big(x_1^{[m_1]}\cdots x_n^{[m_n]} \big),
\label{21.1}
\end{eqnarray}
where $\mathcal M_n$ is the set of all indices $m_1,\ldots,m_n$ with the constraint that $|m_j-m_k|>L$ whenever $j\neq k$ and ${\mathcal M_n}^c$ is the collection of all remaining indices.
`Most' indices are such that all the $m_j$ are apart from each other by at least $L$: The cardinality of $\mathcal M_n$ satsifies 
\begin{equation}
\label{081}
\lim_{M\rightarrow\infty}\frac{|\mathcal M_n|}{M^n} =1
\end{equation}
and $|\mathcal M_n^c| = M^n-|\mathcal M_n|\rightarrow 0$ as $M\rightarrow\infty$. We have
\begin{eqnarray}
\lefteqn{
\frac{1}{M^n}\sum_{\mathcal M_n}  \omega_{\r,M}\big(x_1^{[m_1]}\cdots x_n^{[m_n]} \big)- \omega_\r(x_1)\cdots \omega_\r(x_n)}\nonumber\\
&& = \sum_{\mathcal M_n} \frac{\omega_{\r,M}\big(x_1^{[m_1]}\cdots x_n^{[m_n]} \big)-\omega_\r(x_1)\cdots \omega_\r(x_n)}{M^n} + \big(\frac{|\mathcal M_n|}{M^n}-1\big)\omega_\r(x_1)\cdots \omega_\r(x_n).
\label{082}
\end{eqnarray}
Let $\epsilon>0$. Due to \eqref{corrstate} there is an $M_0>0$ such that if $M>M_0$ then 
$$
\big|\omega_{\r,M}\big(x_1^{[m_1]}\cdots x_n^{[m_n]} \big)-\omega_\r(x_1)\cdots \omega_\r(x_n)\big| <\epsilon,
$$
uniformly in the $m_1,\ldots,m_n$. The absolute value of the sum over $\mathcal M_n$ on the right side of \eqref{082} is thus bounded above by $\epsilon \frac{|\mathcal M_n|}{M^n}$ for $M>M_0$. So as $M\rightarrow\infty$ the upper bound is $\epsilon$ by \eqref{081}, so that this limit is zero. Again by \eqref{081} the second term on the right side of \eqref{081} vanishes as well for $M\rightarrow\infty$. We conclude that 
\begin{equation}
\label{22.1}
\lim_{M\rightarrow\infty} \frac{1}{M^n}\sum_{\mathcal M_n} \omega_{\r,M}\big(x_1^{[m_1]}\cdots x_n^{[m_n]} \big)= \omega_\r(x_1)\cdots \omega_\r(x_n). 
\end{equation}
Next by the estimate \eqref{finitemoment},
$$
\Big| \frac{1}{M^n}\sum_{\mathcal M_n^c}\omega_{\r,M}\big(x_1^{[m_1]}\cdots x_n^{[m_n]} \big)\Big| \le \frac{M^n - |\mathcal M_n|}{M^n}\, C(n).
$$
The left hand side thus vanishes in the limit $M\rightarrow\infty$. Combining this with \eqref{21.1} and \eqref{22.1} yields the result \eqref{20.1}. This completes the proof of Proposition \ref{prop1'}.\hfill $\square$

\subsection{Proof of Proposition \ref{prop:definettidyn}}
\label{sec:definetti}

For bounded $x_j$, we have  ({\it c.f.} \eqref{20.1}) 
\begin{equation}
\label{defi1}
\lim_{M\rightarrow\infty} \omega_{\r,M}\big(\overline X_{M,1}\cdots \overline X_{M,n}\big) = \int_{\mathcal S_\r} d\mu(\omega)\,  \omega(x_1)\cdots \omega(x_n).
\end{equation}
The proof of \eqref{defi1} is as in Proposition \ref{prop1'}, where in the sum over $\mathcal M_n$ in \eqref{21.1} we use the property \eqref{definetti} and the sum over $\mathcal M_n^c$ times $M^{-n}$ vanishes as $M\rightarrow\infty$. The right hand side of \eqref{44} is then $\int_{\mathcal S}d\mu(\omega) G_\omega'(t_{\ell_1})\cdots G_\omega'(t_{\ell_L})\rho_\s G_\omega'(t_{\ell+1})\cdots G_\omega'(t_{\ell_n})$, where $G_\omega'(t) =\omega(v(t))G(t)$. Continuing in the same way as in the proof of Theorem \ref{thm1} we arrive at (compare with \eqref{26}),
\begin{equation}
\label{defin0}
e^{i t H_\s}\rho_\s(t)e^{-i t H_\s} = \int_{\mathcal S_\r} d\mu(\omega)\sum_{n\ge 0}(-i)^n\int_{0\le t_n\le\cdots\le t_1\le t}  [G'_\omega(t_1),[\cdots [G_\omega'(t_n),\rho_\s]\cdots].
\end{equation}
It follows that 
\begin{equation}
\label{defin1}
e^{i tH_\s}\rho_\s(t) e^{-i t H_\s} = \int_{\mathcal S_\r} d\mu(\omega)\,  U'(\omega,t)\rho_\s U'(\omega,t)^*,
\end{equation}
where for $\omega\in\mathcal S_\r$ fixed,
$$
i\partial_t U'(\omega,t) = \omega\big(v(t)\big) G(t) \,  U'(\omega,t), \qquad U'(\omega,0)=\bbbone.
$$
To show that the right side of \eqref{defin1} equals that of \eqref{defin0}, it suffices to note that both quantities solve the same differential equation (obtained upon taking $\partial_t$) and they have the same initial condition for $t=0$ (namely, $\rho_\s$). 
 Thus \eqref{definetti3} holds with $U(\omega,t) = e^{-i t H_\s}U'(\omega,t)$. \hfill $\qed$

\subsection{Proof of Proposition \ref{prop:2a}}
\label{sect:prop2a}

We show \eqref{Cncoherent}. The coherent state $|\alpha\rangle$ of a single mode (oscillator)  is defined by $|\alpha \rangle = W(\alpha) |0 \rangle $, where $| 0 \rangle$ is the vacuum state and the Weyl operator is defined by 
$W(\alpha) = e^{i\frac{1}{\sqrt 2}(\alpha a^* + \bar{\alpha} a)}$ and satisfies the commutation relation 
\begin{equation}
    \varphi(t) W (\alpha) = W (\alpha) \big( \varphi(t) - {\rm Im}(e^{i \Omega_R t} \alpha) \big) .
    \label{comrel_w_f}
\end{equation}
We now obtain a bound of the type \eqref{10} for $\rho_{\r,M}=|\alpha\rangle\langle\alpha|\otimes\cdots\otimes|\alpha\rangle\langle\alpha|$. We estimate,
\begin{align}
    \tr_{\r,M} \big( & (|\alpha \rangle \langle \alpha | )^{\otimes M}  \overline{V}_{\r,M} (t_1) \cdots \overline{V}_{\r,M} (t_n)  \big)  \nonumber \\
    = & M^{-n} \sum_{f: \{1,\dots,n\} \rightarrow \{1,\dots,M\} } \prod_{m=1}^M \langle \alpha | \mathcal{T} \prod_{q \in f^{-1}(m) } \varphi (t_{q}) \alpha \rangle \nonumber \\
    = & M^{-n} \sum_{f: \{1,\dots,n\} \rightarrow \{1,\dots,M\} } \prod_{m=1}^M \sum_{S \subset f^{-1}(m) } \big(\prod_{s' \in S^C } {\rm Im}(-e^{i \Omega_\r t_{s'}} \alpha) \big)  \langle 0 | \mathcal{T} \prod_{s \in S} \varphi (t_s) 0 \rangle \nonumber \\
    = & M^{-n} \sum_{f: \{1,\dots,n\} \rightarrow \{1,\dots,M\} } \prod_{m=1}^M\  \sum_{\substack{S \subseteq f^{-1}(m), \\ |S| \text{ even} }}\  \prod_{s' \in S^C} {\rm Im}(-e^{i \Omega_\r t_{s'}} \alpha )  \cdot \nonumber\\
    & \hspace{5cm} \cdot  \sum_{p \in \mathcal{P}_S} \prod_{(s_1,s_2) \in p  } e^{-i \Omega_\r (t_{s_1}-t_{s_2})}.
\label{mi41}
\end{align}
Here, $\mathcal T$ is the time ordering operator and we sum over all $M^n$ functions $f$ from the set $\{1,\ldots,n\}$ to the set $\{1,\ldots,M\}$. In the second step we use the commutation relation \eqref{comrel_w_f}. In the last step we use Wick's theorem to express the average over a product of field operators by a sum over two-point correlation functions $\langle 0|\varphi(t)\varphi(s)|0\rangle =e^{-i\Omega_\r(t-s)}$. $\mathcal P_S$ is the set of all pairings with indices in $S$. The number of such pairings is $|\mathcal P_S| = |S|!!$, where
\begin{equation}
\label{k!!}
k!! \equiv  \frac{ k!}{2^{k/2} (k/2)!} = (k-1)\cdot(k-3)\cdots 3\cdot 1,\qquad \mbox{$k$ even.}
\end{equation}
We have,
\begin{align}
 \Big|   \sum_{\substack{S \subseteq f^{-1}(m), \\ |S| \text{ even} }} & \ \prod_{s' \in S^C}  {\rm Im}(-e^{i \Omega_\r t_{s'}}\alpha ) \sum_{p \in \mathcal{P}_S} \prod_{(s_1,s_2) \in p  } e^{-i \Omega_\r (t_{s_1}-t_{s_2})} \Big| \nonumber \\
    \leq & \sum_{\substack{S \subseteq f^{-1}(m), \\ |S| \text{ even} }} |\alpha|^{ |f^{-1}(m)|-|S| } \, (|S|!!) \ \leq  \ (1+|\alpha|)^{|f^{-1}(m)|} \ (S_{f,m})!!,
\label{mi42}
\end{align}
where 
\begin{equation}
\label{miSfm}
S_{f,m} = 2 \lfloor | f^{-1}(m) |/2 \rfloor
\end{equation}
is the cardinality of the biggest subset of $f^{-1}(m)$ with an even number of elements. (Here, $\lfloor x\rfloor$ is the floor function; {\it e.g.} $\lfloor\pi\rfloor =3$.) The first factor on the right side of \eqref{mi42} is due to 
$$
\sum_{S \subset f^{-1}(m)} |\alpha|^{ |f^{-1}(m)|-|S| }  =\sum_{k=0}^{|f^{-1}(m)|} {|f^{-1}(m) |\choose k} |\alpha|^{|f^{-1}(m)|-k} = (1+|\alpha|)^{|f^{-1}(m)|}. 
$$
We use \eqref{mi42} in \eqref{mi41},
\begin{align}
\Big|    \tr_{\r,M} \big( & (|\alpha \rangle \langle \alpha | )^{\otimes M}  \overline{V}_{\r,M} (t_1) \cdots \overline{V}_{\r,M} (t_n)  \big) \Big| \nonumber \\
    \le & \ M^{-n} \sum_{f: \{1,\dots,n\} \rightarrow \{1,\dots,M\} } \prod_{m=1}^M \ (1+|\alpha|)^{|f^{-1}(m)|} \ (S_{f,m})!!\, .
    \label{mi45}
\end{align}
We use $\sum_{m=1}^M |f^{-1}(m)|=n$ to estimate the product of the factors containing the $\alpha$ by $(1+|\alpha|)^n$. Observe that for any integer any integers $n_1, n_2$,
\begin{equation}
\label{mi44}
    (2(n_1+n_2))!! \geq (2n_1)!! \cdot (2n_2)!! .
\end{equation}
To see this, we write down the product explicitly and estimate,
\begin{align*}
   (2(n_1+n_2))!! & =  (2n_1 - 1 +2n_2) \cdot (2n_1 - 3 +2n_2) \cdots (3 +2n_2) \cdot (1 +2n_2) \cdot (2n_2)!! \\
  & \geq  (2n_1 - 1 ) \cdot (2n_1 - 3 ) \cdots (3) \cdot (1) \cdot (2n_2)!!\nonumber\\
  & = (2n_1)!! \cdot (2n_2)!!\, .
\end{align*}
It follows from \eqref{miSfm},  \eqref{mi44} and $\sum_{m=1}^M |f^{-1}(m)|=n$ that 
\begin{equation}
\label{mi47}
\prod_{m=1}^M (S_{f,m})!! = \prod_{m=1}^M  (2 \lfloor | f^{-1}(m) |/2 \rfloor)!! \le \Big(2\sum_{m=1}^M\lfloor \tfrac{|f^{-1}(m)|}{2}\rfloor \Big)!! \le \big(2\lfloor n/2\rfloor\big)!!\, .
\end{equation}
Finally, as the number of terms of the sum over the $f$ equals $M^n$, we combine \eqref{mi45} and \eqref{mi47} to arrive at 
\begin{equation}
\Big|    \tr_{\r,M} \big(  (|\alpha \rangle \langle \alpha | )^{\otimes M}  \overline{V}_{\r,M} (t_1) \cdots \overline{V}_{\r,M} (t_n)  \big) \Big| \le (2 \lfloor n/2 \rfloor)!! \ (1+|\alpha|)^n .
\end{equation}
This shows the validity of the expression \eqref{Cncoherent}. \hfill $\qed$

We now indicate how to modify this proof to arrive at Proposition \ref{prop:3a}. Analogously to \eqref{comrel_w_f}, we have the commutation relations  $\varphi(g) W (f) = W (f) ( \varphi(g) -{\rm Im} \langle g | f \rangle )$ and we can replicate exactly the combinatorial argument as in the oscillator case. The two-point correlation functions in Wick's theorem is now $\langle \Omega | \varphi(e^{i\omega t}g) \varphi(e^{i\omega s}g) \Omega \rangle =  \frac{1}{2} \langle g| e^{i \omega (t-s) } g \rangle$, which is bounded in modulus by $\tfrac12 \|g\|^2$.

\subsection{Proof of Proposition \ref{prop:2b}}
\label{sect:prop2b}

As $v_2$ is stationary under the reservoir evolution, we have
\begin{eqnarray}
\lefteqn{
\omega_{\r,M} \big(  \oV_{\r,M}(t_1)\cdots \oV_{\r,M}(t_n)\big) = \frac{1}{M^n}\sum_{m_1=1}^M\cdots\sum_{m_n=1}^M  \omega_{\r,M}\big( v_2^{[m_1]}\cdots v_2^{[m_n]}\big)}\nonumber\\
&&= \frac{1}{M^n} \omega_{\r,M}\Big(\Big(\sum_{m=1}^M v_2^{[m]}\Big)^n\Big)= \frac{\nu^n}{M^n} \omega_{\r,M}\Big(\Big(\sum_{m=1}^M \widehat N_m\Big)^n\Big),
\label{42}
\end{eqnarray}
where $\widehat N_m =(a^\dag a)^{[m]}$ is the number operator acting on the $m$-th oscillator. Using the multinomial expansion for $(\sum_{m=1}^M \widehat N_m )^n$ in \eqref{42}, 
\begin{align}
\omega_{\r,M} \big(  \oV_{\r,M}(t_1)\cdots \oV_{\r,M}(t_n)\big) \nonumber\\
=  \frac{\nu^n}{M^n} & \sum_{\substack{k_1+\cdots +k_M=n\\ k_1,\ldots,k_M \ge 0}} {n \choose k_1,\ldots,k_M} \omega_{\r,M}\Big( (\widehat N_1)^{k_1}\cdots (\widehat N_M)^{k_M}\Big).
\label{43}
\end{align}

Due to \eqref{44.1}, we have $\omega_{\r,M}( (\widehat N_1)^{k_1}\cdots (\widehat N_M)^{k_M})\le c^{k_1+\cdots+k_M}=c^n$ and we obtain from \eqref{43},
\begin{equation}
\omega_{\r,M} \big(  \oV_{\r,M}(t_1)\cdots \oV_{\r,M}(t_n)\big)
\le  \frac{(cv)^n}{M^n}  \sum_{\substack{k_1+\cdots +k_M=n\\ k_1,\ldots,k_M \ge 0}} {n \choose k_1,\ldots,k_M} = (cv)^n.
\label{45}
\end{equation}
It follows that \eqref{10} holds with $C_{\r,n} = (cv)^n$ and $b_\r(t)=1$. \hfill $\qed$

\subsection{Proof of Proposition \ref{prop:3b}}
\label{sect:prop3b}

The number operator of the `site' $m$ is defined by
\begin{equation}
\widehat N_m = \int_{\mathbb R^d} (a^\dag_k a_k)^{[m]} d^dk.
\end{equation}
It commutes with the interaction, $\big[\widehat N_m, (v_2(t))^{[m']}\big] = 0$. Furthermore, we have the well known bound
\begin{equation}
\label{51}
\big\| (v_2(t))^{[m]} (\widehat N_m+1)^{-1}\big\|= \big\|v_2^{[m]} (\widehat N_m+1)^{-1}\big\|\le 2 \|g\|^2,
\end{equation}
where $\|g\|^2=\int_{\mathbb R^d} |g(k)|^2 d^dk$ is the $L^2$ norm of the function $g$. We use \eqref{51} and the fact that $\widehat N_m$ commutes with $(v_2(t))^{[m']}$ to get,
\begin{align}
\Big|\omega_{\r,M}\big(v_2(t_1)^{[m_1]} \cdots v_2(t_n)^{[m_n]}  \big)\Big| \label{52}\\
= \Big|\omega_{\r,M}  \big(w_{m_1,\ldots,m_n}^{1/2} v_2(t_1)^{[m_1]} & (\widehat N_{m_1}+1)^{-1} \cdots v_2(t_n)^{[m_n]} (\widehat N_{m_n}+1)^{-1} w_{m_1,\ldots,m_n}^{1/2} \big)\Big|,\nonumber
\end{align}
where $w_{m_1,\ldots,m_n}= (\widehat N_{m_1}+1)\cdots (\widehat N_{m_n}+1)$. Next, using that for any state $\omega$, any self-adjoint operator $w\ge 0$ and any bounded operator $A$, we have $|\omega(w^{1/2} A w^{1/2})|\le \|A\|\, \omega(w)$, we obtain from \eqref{51}, \eqref{52}, 
\begin{equation}
\Big|\omega_{\r,M}\big(v_2(t_1)^{[m_1]} \cdots v_2(t_n)^{[m_n]}  \big)\Big| \le \big(2
 \|g\|^2\big)^n  \omega_{\r,M}  \big( (\widehat N_{m_1}+1)\cdots (\widehat N_{m_n}+1)\big).
\end{equation}
It follows that 
\begin{align}
\Big|\omega_{\r,M} \big(  \oV_{\r,M}(t_1)\cdots \oV_{\r,M}(t_n)\big)\Big|   \nonumber\\
\le \frac{(2\|g\|^2)^n}{M^n}  
\sum_{m_1=1}^M\cdots & \sum_{m_n=1}^M \omega_{\r,M}  \big( (\widehat N_{m_1}+1)\cdots  (\widehat N_{m_n}+1)\big)\nonumber
\\
=\frac{(2\|g\|^2)^n}{M^n}  
 & \ \omega_{\r,M}  \Big( \big( \sum_{m=1}^M \widehat N_{m}+1\big)^n\Big).
\label{55}
\end{align}
We may now proceed as after \eqref{42} to conclude that if $\omega_{\r,M}$ is the $M$-fold tensor product of $\omega_\r$ satisfying \eqref{44.1.1}, then \eqref{10} holds with $C_{\r,n} = (2c\|g\|^2)^n$ and $b_\r(t)=1$.\hfill $\qed$

\subsection{Proof of Proposition \ref{prop:Stark}} 

We start by noticing that for operators $A,B$ and a normalized state $\psi$,
\begin{equation}
\label{72}
\| A |\psi\rangle\langle\psi| B\|_1 = {\rm Tr} \sqrt{B^*|\psi\rangle\langle\psi| A^*A|\psi\rangle\langle\psi| B} =  \|A\psi\|\, \|B^*\psi\|. 
\end{equation}
We are then led to find a bound on 
\begin{equation}
\label{66.0}
\| G(t_1)\cdots G(t_n) \psi\| =\| e^{i t_1 (-\Delta)} (E \cdot x)  e^{-i t_1 (-\Delta)} \cdots e^{i t_n (-\Delta)} (E \cdot x)  e^{-i t_n (-\Delta)} \psi \|.
\end{equation}
Let us introduce the creation and annihilation operators in the direction $\hat E$ 
\begin{equation}
\label{66}
    a_E = \frac{1}{\sqrt{2}} \,  \hat E \cdot (x+i(-i \nabla_x)), \qquad  a^\dag_E = \frac{1}{\sqrt{2}}\,   \hat E\cdot (x-i(-i \nabla_x)).
\end{equation}
They satisfy the canonical commutation relations $[a_E,a^\dag_E]=1$. We define the field operator 
\begin{equation}
\label{defphi}
\varphi_E(\alpha) = \frac{\alpha  a^\dag_E + \bar{\alpha}a_E}{\sqrt 2}, \qquad \alpha \in \mathbb C. 
\end{equation}
It follows from \eqref{66} that $
\hat E\cdot x = \varphi_E(1)$ and $\hat E\cdot (-i\nabla_x) = \varphi_E(i)$. 
By passing to the Fourier transform, one readily sees that 
$$
e^{i s(-\Delta)} (\hat E\cdot x)e^{-is(-\Delta)} = \hat E\cdot x+2s \hat E\cdot(-i\nabla_x) = \varphi_E(1+2is).
$$
We further define the number operator 
$$
N_E=a^\dag_E a_E
$$
and we have the commutation rules 
\begin{equation}
N_E a_E = a_E (N_E-1)\quad \mbox{and}\quad N_E a^\dag_E = a^\dag_E (N_E+1).
\label{commrel}
\end{equation}
To estimate \eqref{66.0} we express the operators in terms of the fields and insert suitable weights,
\begin{align}
\| G(t_1)\cdots G(t_n) \psi\| =  &  \| \varphi_E (1+2it_1) \dots  \varphi_E (1+2it_n) \psi \| \nonumber\\
=\| \varphi_E (1+2it_1 ) & (N_E+1)^{-\frac{1}{2}} (N_E+1)^{\frac{1}{2}} \varphi_E (1+2it_2) (N_E+2)^{-1} (N_E+2) \cdots \nonumber\\
 \cdots (N_E+(n-1))^{\frac{n-1}{2}} &\varphi_E(1+2i t_n) (N_E+n)^{-\frac{n}{2}} (N_E+n)^{\frac{n}{2}}  \psi \|.
 \label{bnd}
\end{align}
We now need to estimate the norm of the weighted field operators, $(N_E+l)^{\frac{l}{2}} \varphi_E(1+2i t) (N_E+(l+1))^{-\frac{l+1}{2}}$, for $l \in \mathbb{N} \cup \{ 0 \}$. It follows from \eqref{commrel} that
\begin{eqnarray*}
(N_E+l)^{\frac{l}{2}} \varphi_E(1+2i t) &=& \frac{1}{\sqrt 2} (N_E+l)^{\frac{l}{2}} \big( (1+2it) a^\dag_E + (1-2it)a_E\big)\nonumber\\
&=& \frac{1}{\sqrt 2}\big((1+2it) a^\dag_E\,  (N_E+l+1)^{\frac l2}+ (1-2it)a_E (N_E+l-1)^{\frac l2} \big).
\end{eqnarray*}
We combine this with the standard estimate \cite{BR}
$$
\|a_E(N_E+1)^{-1/2}\|\le 1,\quad \|a^\dag_E(N_E+1)^{-1/2}\|\le 1
$$
to obtain,
\begin{equation}
\label{76}
\|(N_E+l)^{\frac l2} \varphi_E(1+2i t) \big(N_E+l+1\big)^{-\frac{l+1}{2}} \| \le \sqrt{2(1+4t^2)}.
\end{equation}
Using the bound \eqref{76} in \eqref{bnd} gives
\begin{equation}
\| G(t_1)\cdots G(t_n) \psi\| \le 
2^{\frac n2} \Big( \prod_{j=1}^n \sqrt{1+4 t_j^2}\, \Big)  \| (N_E+n)^{\frac{n}{2}} \psi\|\le 2^{\frac n2} \Big( \prod_{j=1}^n \sqrt{1+4 t_j^2}\, \Big)  (1+n)^{\frac n2}\| N_E^{\frac n2}\psi\|,  
\label{79}
\end{equation}
where we used 
$\| (N_E+n)^{\frac{n}{2}} \psi\|^2 = \langle \psi , (N_E+n)^n \psi \rangle \leq \langle \psi, (1+n)^n N_E^n \psi \rangle = (1+n)^n \| N_E^{\frac{n}{2}} \psi \|^2$. The bounds \eqref{79} and \eqref{72} give for a normalized $\psi$,
\begin{align}
\label{10.30}
\big \|G(t_{\ell_1})\cdots G({t_{\ell_L}})  |\psi\rangle\langle \psi|  G(t_{\ell_{L+1}}) \cdots G(t_{\ell_n})\big\|_1 \nonumber\\
\le (2(1+n))^{\frac n2} \Big( &\prod_{j=1}^n \sqrt{1+4 t_j^2}\, \Big)
\|N_E^{\frac L2}\psi\| \, \|N_E^{\frac{n-L}{2}}\psi\|.
\end{align}
The bound \eqref{regularity} implies that for all $k\in\mathbb N$,
\begin{equation}
\label{81}
\|N_E^k \psi\|\le c^k.
\end{equation}
Then \eqref{10.30} shows that \eqref{10} holds with  
\begin{equation}
C_{\s,n} = \big(2c(1+n)\big)^{\frac n2},   \qquad  b_\s(t) = \sqrt{1+4t^2}.
\label{82}
\end{equation}
As we are assuming that $C_{\r,n}\le (c')^n$ and $b_\r(t)=1$ locally integrable, the condition \eqref{cond} reads
\begin{equation*}
\sum_{n\ge 0}\frac{\delta^n (1+n)^{\frac n2}}{n!}<\infty,\qquad \delta =  c' \sqrt{2c}  \int_0^t b_\r(s)  \sqrt{1+4s^2}ds.
\end{equation*}
The ratio test shows that the series converges absolutely for all values of the parameters.  Finally, suppose $\rho_\s = \sum_{\alpha\ge 0} p_\alpha |\psi_\alpha\rangle\langle\psi_\alpha|$ such that each $\psi_\alpha$ satisfies \eqref{81}. Then 
\begin{align*}
\big \|G(t_{\ell_1})\cdots G({t_{\ell_L}})  \rho_\s G(t_{\ell_{L+1}}) \cdots G(t_{\ell_n})\big\|_1\nonumber\\
\le \sum_{\alpha\ge 0} p_\alpha \big \|G(t_{\ell_1})\cdots G({t_{\ell_L}}) &  |\psi_\alpha\rangle\langle\psi_\alpha|  G(t_{\ell_{L+1}}) \cdots G(t_{\ell_n})\big\|_1\le C_{\s,n}b_\s(t_1)\cdots b_\s(t_n)
\end{align*}
for the same choice of constants \eqref{82}. This completes the proof of Proposition \ref{prop:Stark}. \hfill $\qed$

\end{document}